\documentclass[aip,apl,amsmath,amssymb,reprint,nofootinbib]{revtex4-1}

\usepackage{graphicx}
\usepackage{dcolumn}
\usepackage{bm}
\usepackage{xcolor} 
\usepackage[utf8]{inputenc}
\usepackage[T1]{fontenc}
\usepackage{mathptmx}
\usepackage{acronym}
\usepackage{siunitx}
\usepackage{comment}
\usepackage[colorinlistoftodos]{todonotes}
\usepackage{mathrsfs}

\usepackage{multirow}

\usepackage{scrextend}

\usepackage{glossaries}
\newacronym{MDM}{MDM}{mode division multiplexing}
\newacronym{SLM}{SLM}{spatial light modulator}
\newacronym{HOM}{HOM}{higher-order mode}
\newacronym{FMF}{FMF}{few-mode fiber}
\newacronym{SBS}{SBS}{stimulated Brillouin scattering}
\newacronym{FBS}{FBS}{forward Brillouin scattering}
\newacronym{BBS}{BBS}{backward Brillouin scattering}
\newacronym{SEM}{SEM}{scanning electron microscope}
\newacronym{SC-PCF}{SC-PCF}{solid-core photonic crystal fiber}
\newacronym{PCF}{PCF}{photonic crystal fiber}
\newacronym{BS}{BS}{beam splitter}
\newacronym{PM}{PM}{power-meter}
\newacronym{FPI}{FPI}{Fabry-Perot interferometer}
\newacronym{SFPI}{SFPI}{scanning Fabry-Perot interferometer}
\newacronym{LIA}{LIA}{lock-in amplifier}
\newacronym{PE}{PE}{photoelastic effect}
\newacronym{MB}{MB}{moving boundary}
\newacronym{VOA}{VOA}{variable optical attenuator}
\newacronym{PD}{PD}{photodetector}

\begin{document}
\preprint{AIP/123-QED}

\title[Intermodal Brillouin scattering in solid-core photonic crystal fibers]{Intermodal Brillouin scattering in solid-core photonic crystal fibers}

\author{Paulo F. Jarschel}
  \email{jarschel@ifi.unicamp.br}
  \affiliation{Photonics Research Center, University of Campinas, Campinas 13083-859, SP, Brazil\looseness=-1}
  \affiliation{Gleb Wataghin Physics Institute, University of Campinas, Campinas 13083-859, SP, Brazil\looseness=-1}
\author{Erick Lamilla}
  \affiliation{Photonics Research Center, University of Campinas, Campinas 13083-859, SP, Brazil\looseness=-1}
  \affiliation{Gleb Wataghin Physics Institute, University of Campinas, Campinas 13083-859, SP, Brazil\looseness=-1}
\author{Yovanny A. V. Espinel}
  \affiliation{Photonics Research Center, University of Campinas, Campinas 13083-859, SP, Brazil\looseness=-1}
  \affiliation{Gleb Wataghin Physics Institute, University of Campinas, Campinas 13083-859, SP, Brazil\looseness=-1}
\author{Ivan Aldaya}
  \affiliation{Campus of São João da Boa Vista, State University of São Paulo, São João da Boa Vista 13876-750, Brazil\looseness=-1}
  \affiliation{Photonics Research Center, University of Campinas, Campinas 13083-859, SP, Brazil\looseness=-1}
\author{Julian L. Pita}
  \affiliation{Photonics Research Center, University of Campinas, Campinas 13083-859, SP, Brazil\looseness=-1}
  \affiliation{Gleb Wataghin Physics Institute, University of Campinas, Campinas 13083-859, SP, Brazil\looseness=-1}
  \affiliation{School of Electrical Engineering, University of Campinas, SP, Campinas 13083-852, Brazil\looseness=-1}
\author{Andres Gil-Molina}
  \affiliation{Gleb Wataghin Physics Institute, University of Campinas, Campinas 13083-859, SP, Brazil\looseness=-1}
  \affiliation{School of Electrical Engineering, University of Campinas, SP, Campinas 13083-852, Brazil\looseness=-1}
  
  

\author{Gustavo S. Wiederhecker}
  \affiliation{Photonics Research Center, University of Campinas, Campinas 13083-859, SP, Brazil\looseness=-1}
  \affiliation{Gleb Wataghin Physics Institute, University of Campinas, Campinas 13083-859, SP, Brazil\looseness=-1}
\author{Paulo Dainese}
  \affiliation{Photonics Research Center, University of Campinas, Campinas 13083-859, SP, Brazil\looseness=-1}
  \affiliation{Gleb Wataghin Physics Institute, University of Campinas, Campinas 13083-859, SP, Brazil\looseness=-1}
  \affiliation{Corning Research \& Development Corporation, One Science Drive, Corning, New York 14830, USA\looseness=-1}

\date{\today}


\begin{abstract}
We investigate intermodal forward Brillouin scattering in a solid-core \gls{PCF}, demonstrating efficient power conversion between the HE$_{11}$ and HE$_{21}$ modes, with a maximum gain coefficient of 21.4~$\textrm{W}^{-1}\textrm{km}^{-1}$. By exploring mechanical modes of different symmetries, we observe both polarization-dependent and polarization-independent intermodal Brillouin interaction. Finally, we discuss the role of squeeze film air damping and leakage mechanisms, ultimately critical to the engineering of \gls{PCF} structures with enhanced interaction between high order optical modes through flexural mechanical modes.
\end{abstract}

\newcommand\blfootnote[1]{
  \begingroup
  \renewcommand\thefootnote{}\footnote{#1}
  \addtocounter{footnote}{-1}
  \endgroup
}

\maketitle


\section{\label{section:intro}Introduction:}
\blfootnote{\begin{addmargin}[-0.1em]{0em}Paulo F. Jarschel and Erick Lamilla contributed equally to this work.\end{addmargin}}
Applications exploring optical waveguides and cavities supporting multiple spatial modes have greatly expanded in recent years.
Albeit previously considered as an impairment for optical communications, multimode systems regained attention as the basis of \gls{MDM} \cite{richardson2013space, li2014space}, and can substantially enhance the capabilities in many other applications such as sensing \cite{li2015few, weng2015single, wang_few-mode_2017,murray_speckle-based_2019}, particle manipulation \cite{hsu_manipulation_2013, parker2020optical, yan2013guiding}, and nonlinear optical devices as for example in frequency comb generation in multimode ring resonators\cite{ji2020exploiting} and non-reciprocal devices based on Brillouin scattering in multimode  waveguides\cite{kittlaus2018, otterstrom2019}.
Excitation of different spatial modes can be performed using various approaches such as electronically-addressable \glspl{SLM}~\cite{koebele_two_2011, labroille2014efficient, forbes_creation_2016, fontaine2019laguerre}, photonic lanterns\cite{leon2014mode, velazquez2018scaling}, and integrated mode combiners/multiplexers~\cite{solehmainen2006adiabatic}.  
However, interaction between propagating modes is more difficult to achieve, despite being a crucial functionality for all-optical systems, as in switching, mode conversion, and optical isolators.
Nonlinear optical effects offer a path to enable and control intermodal interactions, for example using Kerr-induced long period gratings to perform all-optical mode conversion~\cite{andermahr_optically_2010}.
\Gls{SBS} is another nonlinear mechanism that can be explored for intermodal interaction, with unique properties~\cite{wiederhecker_brillouin_2019, russell_theory_1991, song_intermodal_2013, song_characterization_2013, li_characterization_2013}.
Besides enabling direct power exchange between different spatial modes, \Gls{SBS} can be explored as a mode-selective isolator or mode-selective \gls{VOA}\cite{huang_complete_2011, kang_reconfigurable_2011}, and has recently been explored as the basis for non-reciprocal devices\cite{kittlaus2018, otterstrom2019}. 

Interaction through Brillouin scattering can occur between modes that are co-propagating or counter-propagating, respectively referred as \gls{FBS} or \gls{BBS}~\cite{wiederhecker_brillouin_2019}.
In integrated silicon waveguides, high intermodal Brillouin gain has been recently reported\cite{kittlaus_-chip_2017}. In fibers, the first demonstration of stimulated intermodal \gls{FBS} was obtained in all-solid fibers~\cite{russell_experimental_1990, russell_theory_1991}, with relatively low gain. Efficient experimental demonstrations were only obtained in \gls{BBS} configuration, using  \glspl{FMF}~\cite{song_intermodal_2013, song_characterization_2013, li_characterization_2013}.
As an alternative, solid-core \glspl{PCF} offer enhanced optomechanical interactions due to the greater flexibility to engineer both optical and mechanical modal properties~\cite{dainese_stimulated_2006, dainese_raman-like_2006, beugnot2007guided, kang_all-optical_2010}, and is the basis for the results presented here. In this paper, we provide a comprehensive experimental and theoretical analysis of intermodal \gls{FBS} in a \gls{PCF}.
By exploring different supported mechanical modes, we show that both polarization-independent and polarization-dependent interaction can be implemented, and demonstrate \textit{forward} intermodal Brillouin gain comparable to \textit{backward} Brillouin-based mode conversion in \glspl{FMF}.
Finally, we investigate the fundamental limitations to conversion efficiency imposed by optomechanical coupling and by different forms of mechanical dissipation.


\begin{figure*}[!t]
\includegraphics{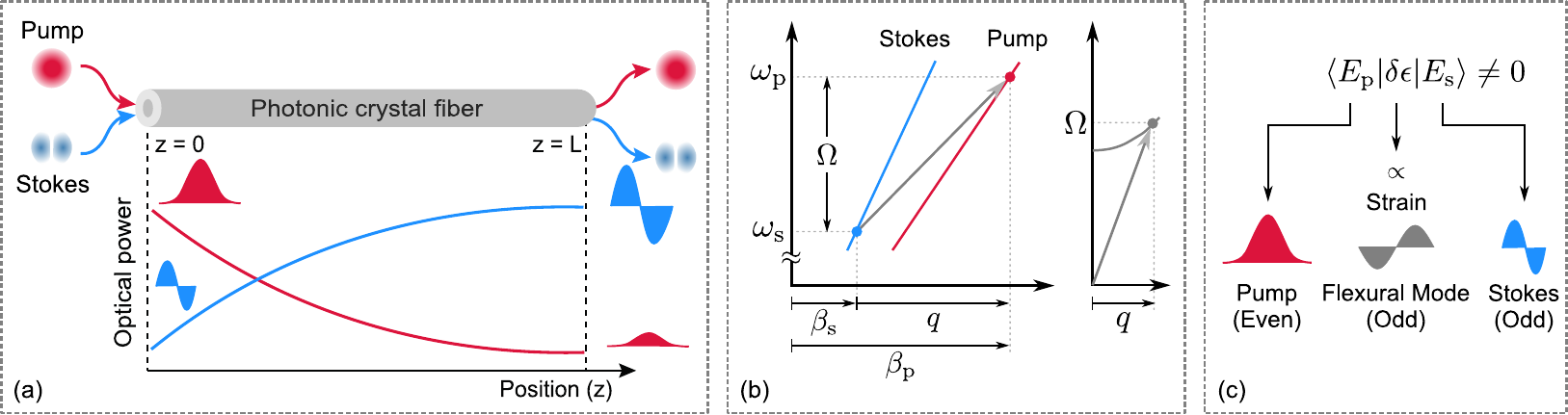}
\caption{\label{Fig:concept}(a) Schematic of energy transfer from the fundamental (pump) to a high-order anti-symmetric mode (Stokes), as they propagate along the fiber. (b) Optical (left) and mechanical (right) dispersion diagrams. Solid blue and red lines represent dispersion relations for pump and Stokes modes, respectively. The grey arrow corresponds to the inter-modal scattering. Phase-matching condition is also sketched at the bottom of the optical dispersion diagram. (c) Non-zero spatial overlap imposes a symmetry selection rule: symmetric fundamental mode and anti-symmetric high-order mode interact through an anti-symmetric mechanical flexural mode.}
\end{figure*}

\section{\label{section:concept}Phase-matching and symmetry considerations}

In intermodal \gls{FBS}, depicted in Fig.~\ref{Fig:concept}(a), optical forces generated by the beating of two optical modes selectively excite mechanical modes in the fiber, which in turn induce power exchange between the optical modes as they propagate.
In particular, we investigate the interaction between the fundamental HE$_{11}$ mode and one of the high-order HE$_{21}$ modes, employed as pump and probe (Stokes line), respectively.
Efficient interaction occurs when both energy and momentum are conserved, i.e. $\omega_p=\omega_s+\Omega$ and $\beta_p=\beta_s+q$, where $\Omega$, $\omega_p$, and $\omega_s$ are respectively the mechanical, pump, and Stokes angular frequencies, and $q$, $\beta_p$, and $\beta_s$ are the corresponding propagation constants (Fig.~\ref{Fig:concept}(b)).
In practice, these conditions yield a process that is highly frequency selective, as the pump-Stokes frequency detuning must match the frequency of a given mechanical mode within the mechanical resonance linewidth, typically in the order of MHz. This property enables Brillouin-based devices to be highly wavelength selective. Another necessary condition is imposed by spatial symmetry selection rules. Following conventional mode coupling notation, the optomechanical coupling coefficient is proportional to the spatial overlap of the interacting fields~\cite{wiederhecker_brillouin_2019}:

\begin{equation}
\label{overlapIntegral}
\langle \Vec{E}_{\text{p}}\mid\delta\Vec{\epsilon}\mid \Vec{E}_{\text{s}}\rangle=\int_S\Vec{E}_p^{*}\cdot\bm\delta \epsilon^{*}\cdot\Vec{E}_s\mbox{ }dS\ne 0,
\end{equation}

\noindent where $S$ is the fiber cross-section, $\Vec{E}_p^{*}$ and $\Vec{E}_s$ are the pump and Stokes electric field profiles, and $\bm \delta\epsilon$ is the tensor of the permittivity perturbation induced by the mechanical mode.
In our particular case, $E_p$ and $E_S$ exhibit even and odd reflection symmetry, respectively.
As a result, the mechanical modes must induce a perturbation with odd symmetry to yield a non-zero overlap integral.
This is satisfied for example by a flexural mechanical mode, as illustrated in Fig.~\ref{Fig:concept}(c).
In the particular case of \glspl{PCF}, several mechanical modes satisfy both phase-matching and spatial overlap selection rule, opening possibilities to explore novel mode-conversion functionalities in a wide range of discrete frequencies with or without polarization selectivity.

\section{\label{section:setup}Experimental setup}

\begin{figure}
\includegraphics[scale=1.0]{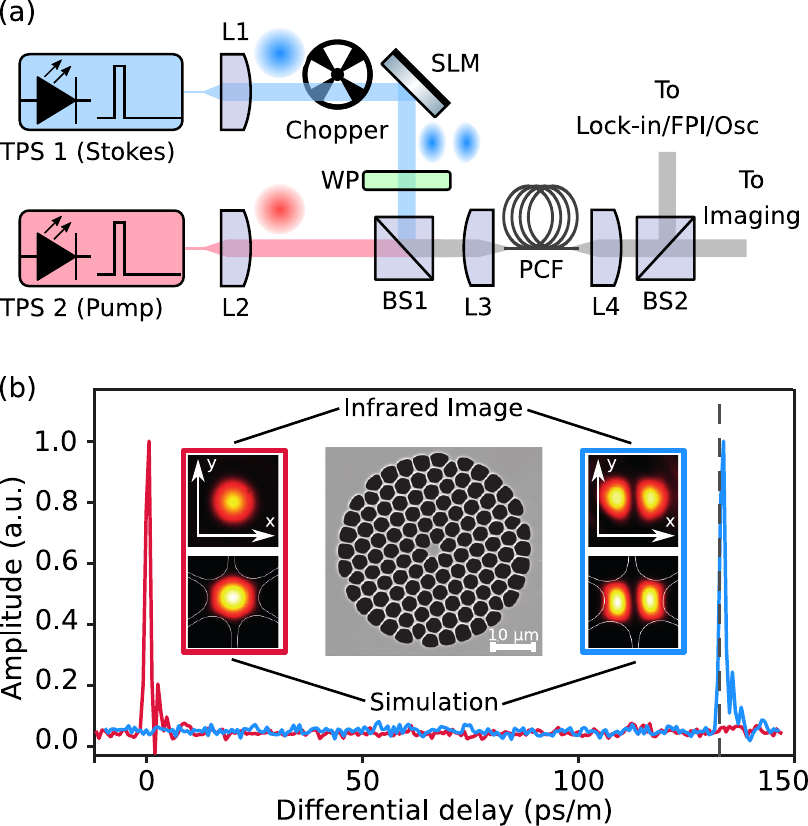}
\caption{\label{Fig:setup}(a) Experimental setup for inter-modal power conversion characterization in \gls{PCF}s. TPS: tunable pulsed source; L: lens; SLM: spatial light modulator; WP: wave plate; BS: beam splitter; FPI: scanning Fabry-Perot interferometer. L3 and L4 are 50x microscope objective lenses.  (b) Time series showing the differential delay corresponding to the HE$_{11}$ and HE$_{21a}$ modes. The dashed line corresponds to the position of the differential delay for the high-order mode, obtained by finite element method simulations. Experimental (top) and simulated (bottom) beam profiles for each scenario and \Gls{SEM} image of the \gls{PCF} under study are shown in the insets.}
\end{figure}

The setup used in our experiments is illustrated in Fig.~\ref{Fig:setup}.
On the launch side, both pump and Stokes lasers propagate in free-space.
The Stokes beam reflects off a \gls{SLM}, which can dynamically transform the phase profile of an incident beam, resulting in the excitation of one of the supported high-order modes (See Supplementary Material S2.A).
A half-wave plate is used to adjust the Stokes polarization to match that of one of the HE$_{21}$ modes, which are roughly linearly polarized.
The pump polarization is adjusted to be either parallel or orthogonal to Stokes (referred to as x and y axis from here on).
Both beams are then combined and launched into the fiber.

Low repetition rate pulses are used for both pump and Stokes (generated using external modulators) in order avoid power damages to the SLM. Such pulsed scheme also helps identifying the optical modes excited through their group delay differences (see Supplementary Material section S2.A for details). Initially, in order to optimize the \gls{SLM} phase masks and characterize the excitation of various guided optical modes, we blocked off the pump beam and used short 35~ps pulses (at a repetition rate of 155~MHz) for the Stokes signal.
The output of the \gls{PCF} was split to enable simultaneous imaging of the beam via an InGaAs camera, and detect the output pulses with a 20~GHz photodiode.
Fig.~\ref{Fig:setup}(b) shows the output pulses, alongside with the output beam profiles for a blank SLM phase mask (red) and a phase profile optimized to excite one of the HE\textsubscript{21} modes (blue).
In each case, the output trace shows a single pulse, either a fast pulse corresponding to the fundamental mode or a delayed pulse, which in this case corresponds to the HE\textsubscript{21a} mode.
The output beam shows clean and well defined profiles, corresponding to the respective simulated mode shapes.
A micrograph of the \gls{PCF} used in our experiments is shown in Fig.~\ref{Fig:setup}(b), from which we measured the following parameters: 3.35$\pm$0.05~\si{\micro\metre} solid-core diameter, 4.0$\pm$0.2~\si{\micro\metre} cladding pitch and 4.0$\pm$0.1~\si{\micro\metre} cladding hole diameter.
Both the differential delay and mode shape results compare well with simulations performed in Comsol Multiphysics (dashed lines and bottom inset images in Fig.~\ref{Fig:setup}b), employing the actual cross-section extracted from the \gls{SEM} image.
Using the same short-pulse modulation, we also verified that the pump arm excites the fundamental mode with a high purity (no other modes are detected in time domain traces).

Once the mode excitation is optimized, we characterized the propagation and coupling losses for both HE\textsubscript{11} and HE\textsubscript{21a} modes using the cutback method. For a fiber length of 30~m we obtained a propagation loss of 0.04~dB/m and coupling loss of 1.5~dB for HE\textsubscript{11}, while for HE\textsubscript{21a}, propagation and coupling losses were determined to be 0.15~dB/m and 4.5~dB, respectively.
Although it is certainly possible to optimize the mode-launching scheme to minimize coupling losses in a practical device, it is not critical here for the purpose of demonstrating proof of principle.

To perform the \gls{FBS} experiments, we switched to longer 80-ns square pulses at a repetition rate of 250~kHz, and each signal was individually amplified by erbium-doped fiber amplifiers, reaching peak powers of up to 2~W while preserving a relative low average power.
The Stokes laser was kept at a constant operating wavelength, while the pump frequency was swept so that different mechanical modes could be excited.
The pump-Stokes frequency detuning was continuously monitored using an electrical spectrum analyzer.
At the output of the fiber, the beam was split and one arm is monitored with an InGaAs camera while the other is directed to two possible detection schemes, selected by a flip mirror.
In the first scheme, we perform a broadband mechanical spectroscopy by sweeping the pump-Stokes frequency detuning over a wide range. Here, a novel detection scheme using a \gls{LIA} is implemented to detect the mechanical resonances signatures. In small gain-regime, the magnitude of the \gls{LIA} signal is linearly proportional to the Brillouin gain spectrum (details regarding this approach can be found in Section S2.B of the Supplementary Material).
While this technique is fast and does not require tunable narrow-band optical filters, it cannot discriminate between the Stokes gain or pump depletion, since the chopper modulation is transferred from Stokes to pump due to the Brillouin interaction along the fiber. We therefore use a second detection scheme once a certain mechanical resonance is identified in the \gls{LIA} spectrum. In this second, a scanning \gls{FPI} (7.5~MHz linewidth, 1.5~GHz free-spectral range) is employed to individually measure the pump and Stokes powers at the output of the fiber, and the energy transfer between the two signals can be quantitatively characterized. 

\begin{figure*}[hbt!]
\includegraphics[scale=0.9]{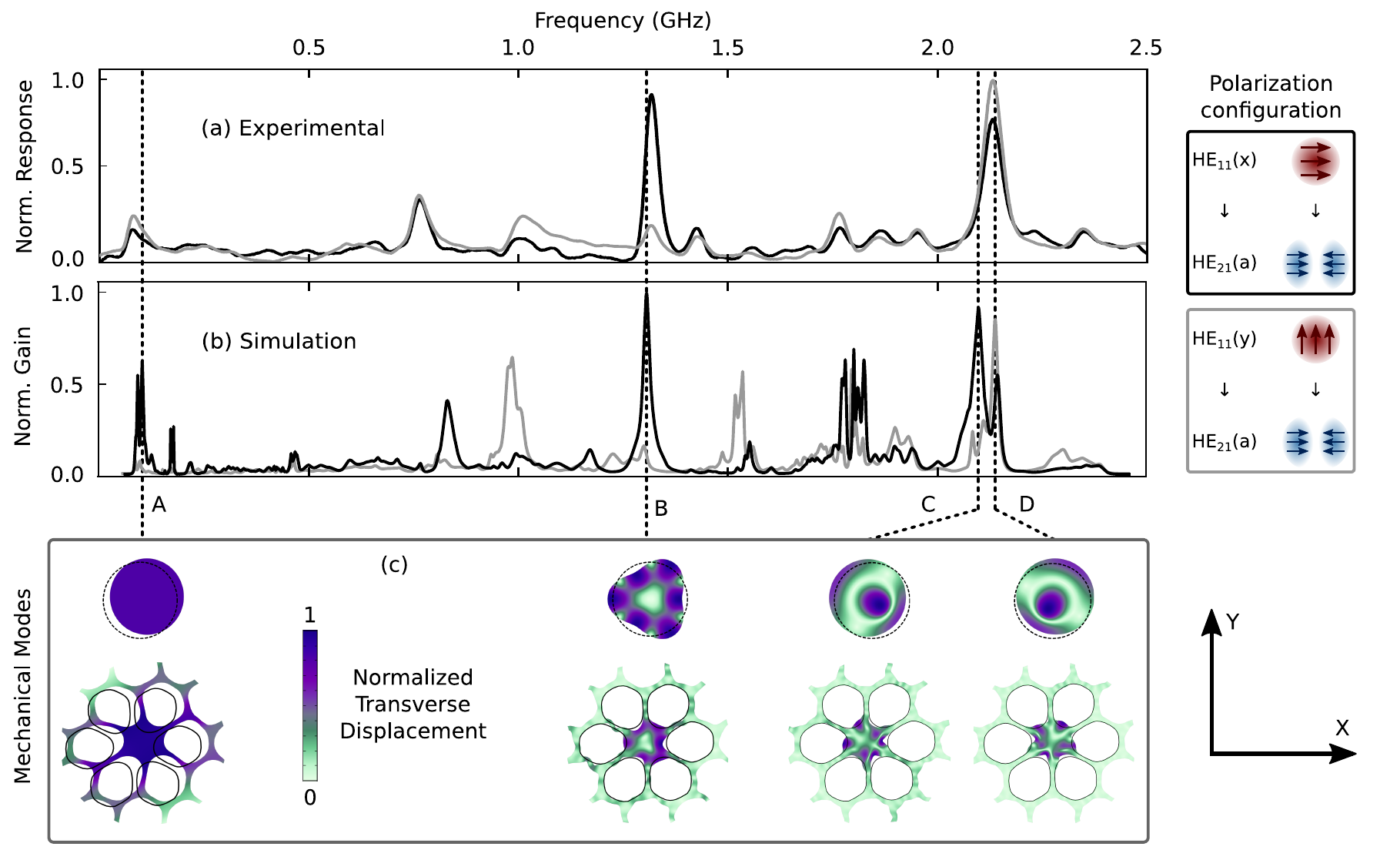}
\caption{\label{Fig:lockin} (a) Experimental lock-in amplifier response (for 1~W of input power for each signal) and (b) simulated Brillouin gain spectra (normalized by the maximum peak), considering energy transfer between the HE$_{11}$ and HE$_{21a}$ modes for parallel (black) and crossed (gray) polarizations, as schematically suggested in the rightmost inset. The mechanical mode profiles for the selected A, B, C, and D peaks are shown in the bottom panel for both the realistic \gls{PCF} model and a silica rod, with the same radius as the fiber core. The color scale represents the intensity of the transverse displacement in the cross section of the PCF.}
\end{figure*}

\section{\label{section:results} Results}

\subsection{Intermodal FBS spectrum}

Using the \gls{LIA} detection method, we first obtained the intermodal Brillouin spectrum shown in Fig.~\ref{Fig:lockin} for parallel and orthogonal pump-signal polarizations.
Multiple mechanical resonances are observed, with two clearly dominant peaks centered at 1.30~\si{\giga\hertz} and 2.13~\si{\giga\hertz}, and weaker resonances at 90~\si{\mega\hertz}, 750~\si{\mega\hertz}, and around 1.0~\si{\giga\hertz} and 1.75~\si{\giga\hertz}.
No other significant resonances were observed between 2.5~\si{\giga\hertz} and 10.0\si{\giga\hertz}.
To better understand the features in the experimental spectrum and identify the mechanical modes responsible for each peak, we simulated the Brillouin gain for all modes supported by the \gls{PCF} structure in the frequency range of interest.
Details on the simulation can be found in Section S1 of the Supplementary Material.
For the calculated Brillouin gain spectrum, the vertical axis represents Brillouin gain (normalized to the highest peak).
Overall, the simulated spectrum semi-quantitatively explains the most important features observed in the experiment.
Particularly, it is possible to identify the mechanical modes with highest gain and obtain the relative strengths of the strongest peaks.
In addition, it gives insight on the observed polarization dependence.
We now discuss these features in more detail.

In order to identify the mechanical modes related to the dominant peaks, we include simulated displacement profiles as insets in Fig.~\ref{Fig:lockin}.
These modes were calculated for a full \gls{PCF} structure (obtained from the SEM profile) and compared to the corresponding modes in a suspended rod with the same core diameter.
As expected, the flexural nature of these modes can be clearly observed, and moreover, the field profiles remarkably resemble those in a simple rod: all peaks correlate to flexural modes, being A the fundamental, and B, C, and D higher-order modes, with displacement nodes within the core region.
Another important aspect is the 2-fold and 3-fold symmetry of these modes, which is critical to understand the polarization dependence observed both in the experiments and simulations.
Interestingly, the strongest mechanical resonances (peak B at 1.30~\si{\giga\hertz} and C/D at 2.13~\si{\giga\hertz}) present significantly different polarization dependency.
On one hand, at 2.13~GHz we observe strong power exchange for both parallel and orthogonal polarization, while at 1.30~GHz only when pump and Stokes have orthogonal polarization. In other words, by selecting the frequency detuning, it is possible to create a device that is either dependent or virtually independent of the polarization.

The physical mechanism behind this observation is quite unique, and not expected based on the simple rod analogy (where the peak at 1.3~GHz is polarization independent, contrary to our observation in a \gls{PCF}). In a rod, there are two degenerate modes with a 6-fold symmetry displacement profile at the 1.3~GHz region, rotated by 30 degrees relative to each other (one orientation is shown as mode B in Fig. \ref{Fig:lockin}). 
For parallel or orthogonal pump-probe polarization, only one or the other of these two  orientations is excited, with however identical Brillouin gain (explaining why in a rod this peak is polarization independent). In a \gls{PCF}, the picture changes. Even though both orientations of this mechanical mode are still present, one of them couples with the photonic crystal cladding much more strongly than the other. This hybridization with the cladding modes simply means that the mechanical energy for that particular orientation is no longer concentrated in the core, which leads to reduced its overlap with the optical mode (thus lower Brillouin gain). The orientation that hybridizes with the cladding is exactly the one that would in principle couple orthogonal pump-probe polarizations, but, as observed experimentally and confirmed in the simulation, it does so very weakly. Physically, it is quite clear why one orientation couples to the cladding more strongly than the other. The 6-fold symmetry of these two modes coincides with the 6-fold symmetry of the photonic crystal cladding. This means that the mode whose maximum displacement lobes coincide with the glass webs hybridizes strongly, while the other, whose displacement nodes are aligned to the glass webs, does not.



\begin{figure*}[htb!]
\includegraphics[scale=0.91]{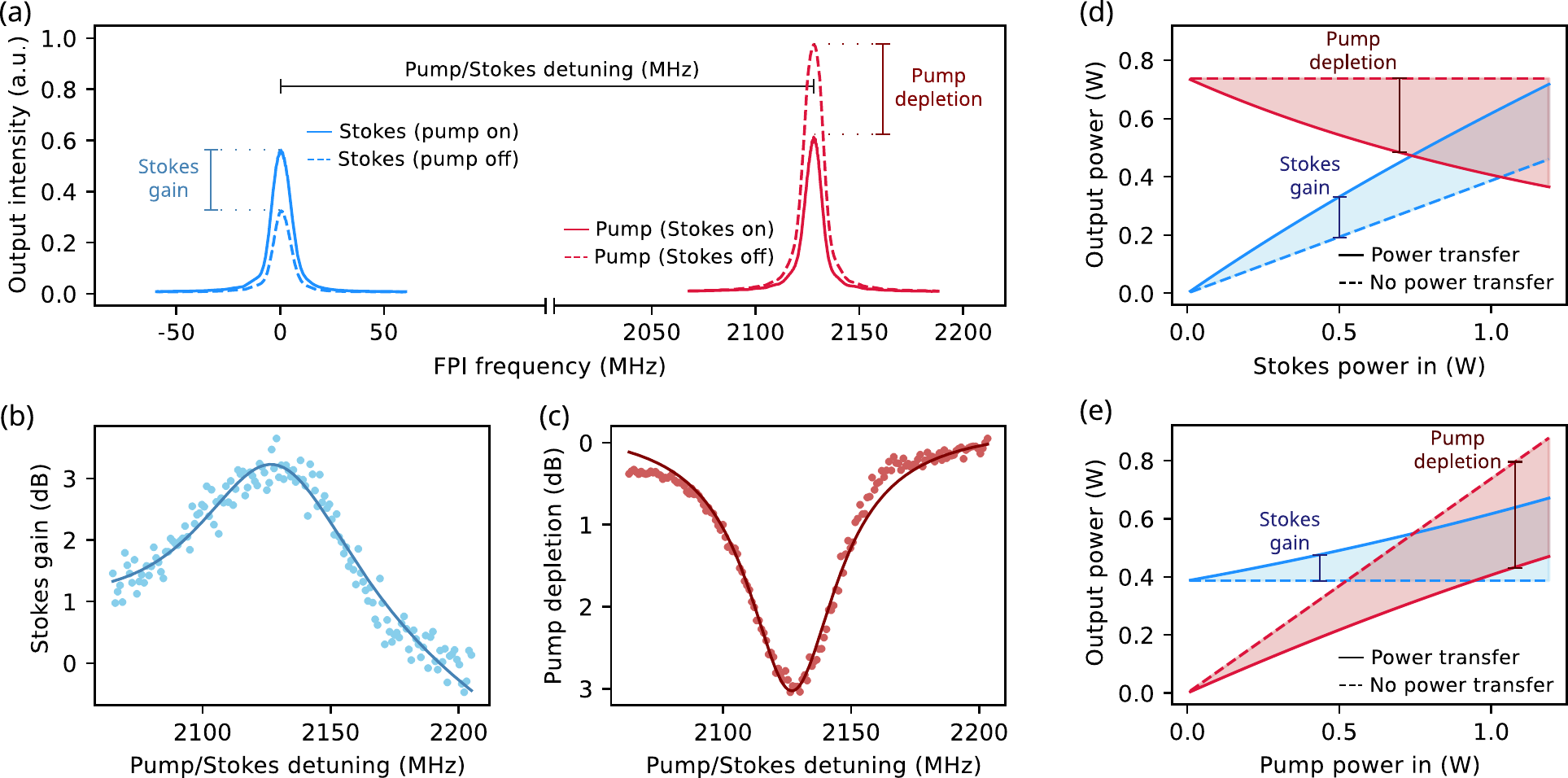}
\caption{\label{Fig:fp1} (a) Examples of measured spectra using the \gls{FPI} scheme. (b, c) Stokes on-off gain and pump on-off depletion versus detuning. (d,e) Measured Stokes (blue) and pump (red) output power levels as a function of Stokes and pump input power. Dashed curves correspond to output power in absence of Brillouin interaction (either pump or Stokes turned off), and solid curve correspond to both lasers interacting.}
\end{figure*}

We now turn to the mechanical modes giving rise to the polarization independent peak at 2.13 GHz.
At about this same frequency, a rod supports two degenerate modes with a 2-fold symmetry, one rotated by 90 degrees relative to the other.
Again, one mode couples parallel polarization, and the other orthogonal. 
Differently than the previous 6-fold mechanical modes, none of the 2-fold orientations hybridizes strongly with the cladding.
Physically, the mismatch in symmetry (2-fold vs. 6 fold) reduces the hybridization of the core modes with the photonic crystal cladding, and one can see in Fig. \ref{Fig:lockin}, that the mechanical energy is mostly concentrated in the core for both modes C and D.
As a result, in the \gls{PCF}, the orientation in C strongly couples parallel polarization, while the rotated mode in D couples orthogonal polarization, resembling the behaviour in a rod. This is also confirmed by numerical simulations (note however that these modes are no longer degenerate, due to slight mechanical distortion in the structure).   

It is worth pointing out that in some of the peaks, one can observe in both experiment and simulated spectra that their shapes deviate from a typical Lorentzian.
This is because these peaks are formed from a cluster of several mechanical modes around their central frequency, due to coupling between core and cladding vibrations. 
Obviously, the pulsed excitation character convolved with the averaging in the LIA signal (which integrates the signal as the pump frequency is swept) effectively broadens the observed spectrum.
Having said that, convolving the simulated spectrum with a filter equivalent to the experimental averaging ($\sim 15$~MHz bandwidth), is not sufficient to fully explain the experiment (see for example a single peak at 2.13~GHz in the experiment while the simulation even after convolution shows two separate peaks). We point out that other possible reasons for this disagreement might be a mismatch between the actual and simulated fiber structure or fluctuations along the fiber length.
Another aspect that upon careful observation might seem surprising is that the fundamental flexural mode at 90~MHz is not the dominant Brillouin peak.
In a rod, one can show that the equivalent mode indeed exhibits the strongest Brillouin gain, and that is not the case in the \gls{PCF} studied here, as observed experimentally and confirmed by the simulations.
We will return to this point when we investigate in more detail the physical process dominating the strength of the optomechanical coupling as well as the damping mechanisms of the mechanical modes.

\subsection{Intermodal FBS energy transfer}

To quantitatively analyze the intermodal energy transfer at the most prominent Brillouin peaks, we performed a fine frequency scan and used the \gls{FPI} detection scheme that alllows discrimination between pump and Stokes signals.
Initially, both have a fixed frequency separated by 2.13~GHz, corresponding to one of the Brillouin peaks in Fig.~\ref{Fig:lockin}, and the \gls{FPI} is scanned to measure the output power of each signal separately.
The results are presented in Fig.~\ref{Fig:fp1}(a) for three conditions: (i) both pump and Stokes on, (ii) only Stokes on, and (iii) only pump on.
We used 1~W of input peak power for the Stokes and 2~W for the pump.
Clearly, when both signals are present, a strong depletion of the pump is accompanied by a Stokes gain, as a result of the intermodal \gls{FBS} process.
A convenient way to quantify the energy transfer is the on/off gain, defined as the ratio of the Stokes output power levels with pump on and off.
Analogously, the pump depletion is the ratio of the pump output power measured when the Stokes signal is turned on and off.
The curves from Figs.~\ref{Fig:fp1}(b, c) show the measured on/off Stokes gain and pump depletion for a narrow sweep around the 2.13~GHz frequency detuning.
As the frequency separation approaches the Brillouin resonance, gain and depletion are maximized, reaching about 3~dB peak values.
The solid lines represent numerical fitting that resulted in a Lorentzian curve with a 45~MHz FWHM.
Performing a deconvolution of the signal with an effective Lorentzian accounting for the pulsed source and the \gls{FPI} linewidth, we estimate a Brillouin linewidth of 42~MHz.

Fig.~\ref{Fig:fp1}(d) shows the evolution of the output power as the input Stokes power increases, while keeping pump input level constant.
Clearly, pump depletion increases with Stokes power, and at the levels experimentally available, we did not reach complete pump depletion.
Similarly, Fig.~\ref{Fig:fp1}(e) shows the output when the pump input power varies and Stokes is kept constant. Note that in both (d) and (e), the drop in pump power is not identical to the increase in Stokes power due to the different propagation losses of each mode.

To extract the gain coefficient, we can numerically solve the coupled equations in \gls{FBS} process and fit the experimental data (See section S2.C of the Supplementary Material for details).
We applied this procedure for three relevant Brillouin peaks by setting the corresponding pump-Stokes frequency detuning, for parallel and orthogonal pump/Stokes polarization conditions.
The results are summarized in Table~\ref{tab:table1}.

\begin{table}
\caption{\label{tab:table1}Gain coefficient values for the major observable Brillouin peaks, for both polarization conditions.}
\begin{ruledtabular}
\begin{tabular}{ccc}
Frequency & Gain (parallel) & Gain (orthogonal)\\
(MHz) & (W$^{-1}$km$^{-1}$) & (W$^{-1}$km$^{-1}$)\\
\hline
90 & $6.1\pm0.1$ & $2.6\pm0.1$\\
1310 & $19.9\pm0.2$ & $3.9\pm0.1$\\
2130 & $19.2\pm0.2$ & $21.4\pm0.3$\\
\end{tabular}
\end{ruledtabular}
\end{table}

\section{\label{section:discussion}Discussion and conclusions}

From the experimental results listed on Table \ref{tab:table1}, the highest gain coefficients are 21.4~$\textrm{W}^{-1}\textrm{km}^{-1}$ for a detuning of 2.13~GHz, and 19.9~$\textrm{W}^{-1}\textrm{km}^{-1}$ for 1.31~GHz.
In fact, this gain coefficient obtained in \textit{forward} configuration in \gls{PCF} is on the same order as observed in \textit{backward} intermodal Brillouin scattering in \glspl{FMF} \cite{song_characterization_2013}. This is quite remarkable because again, in all-solid fibers, forward Brillouin scattering involve mechanical modes that are distributed throughout the cladding with very little overlap with the optical modes. In \gls{PCF}, transverse mechanical confinement enables high gain in forward configuration. 
Having said that, a more complete picture is provided here to better understand the physical limits and possible enhancement directions.

\begin{figure}[htb!]
\includegraphics[scale=0.8]{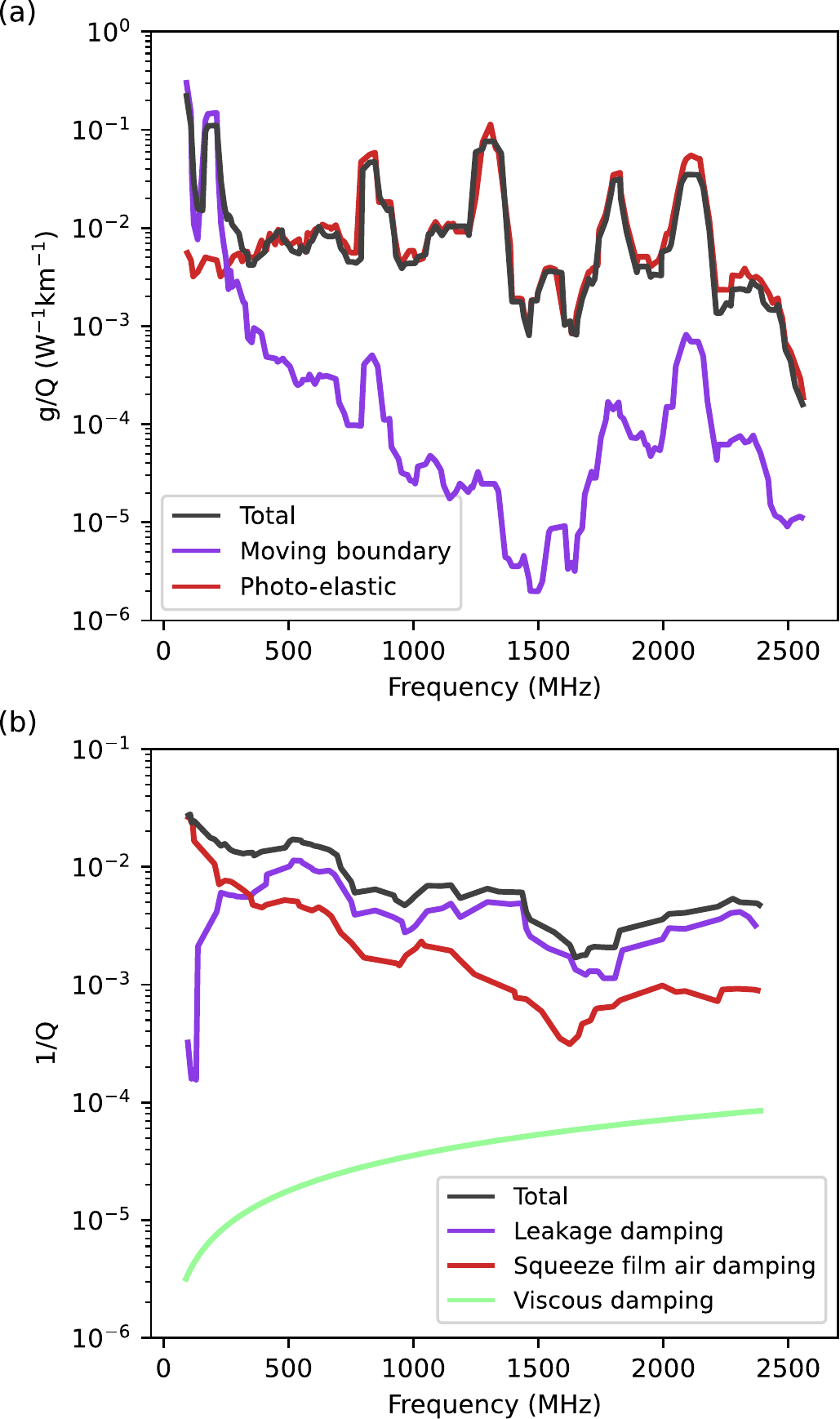}
\caption{\label{Fig:qfactors} (a) Calculated contributions to the net Brillouin gain (g/Q) from moving boundary and photo-elastic effects. (b) Calculated contributions from different mechanisms to the total mechanical damping, represented as the inverse of the quality factor ($Q^{-1}$). Solid lines represent the envelope of the respective raw data, which is available in the supplementary material.}
\end{figure}

Fundamentally, the maximum Brillouin gain for a given resonance depends on the strength of the optomechanical coupling and on the linewidth of that particular resonance.
It is therefore useful to separate the discussion of the optomechanical coupling strength from the damping mechanisms of the mechanical modes.
In Fig.~\ref{Fig:qfactors}, for all mechanical modes we plot separately $g/Q$, which represents the optomechanical coupling strength, and the inverse of $Q$, which represents the damping strength. It is more convenient to look at $Q^{-1}$ as we can directly add contributions from different damping mechanisms (i.e. $Q^{-1} = \sum_i Q^{-1}_i$). The number of mechanical modes in the complete structure is quite large, and thus leads to very dense curves of the calculated coefficients. To help visualize the general trend, the results shown in this figure represent the envelop curves (highest $g/Q$ and lowest $1/Q$), while the raw data for all modes can be found in the supplementary materials.

From the $g/Q$ curves in Fig.\ref{Fig:qfactors}(a), we can see that the lower frequency flexural modes indeed exhibit larger coupling strength (0.3~$\textrm{W}^{-1}\textrm{km}^{-1}$ for 90~MHz) than higher frequencies modes (0.04~$\textrm{W}^{-1}\textrm{km}^{-1}$ for 2.13~GHz), as expected from the suspended rod model.
The results in Fig.~\ref{Fig:qfactors}(a) include contributions from moving-boundary and elasto-optic effects, which can reinforce or counter-act each other~\cite{wiederhecker_brillouin_2019, florez2016brillouin}.
We can see that, for the fundamental flexural modes, large optomechanical coupling is due to the moving-boundary contribution at the glass-air interface, a mechanism that is irrelevant in solid fibers given their small core-clad index constrast.
For higher-order flexural modes, the elasto-optic mechanism dominates with overall lower $g/Q$ values.
Despite stronger optomechanical coupling, our experiment shows that these low frequency flexural modes do not dominate the Brillouin spectrum. This can be explained by evaluating the damping mechanisms shown in Fig.~\ref{Fig:qfactors}(b). Here, the lower frequency modes suffer the strongest damping (poorer quality factors), overriding their high optomechanical coupling.
In this analysis, we considered three loss mechanisms for the mechanical modes: viscosity, leakage, and squeezed film air damping.
Viscous damping scales quadratically with frequency ($Q\propto1/\omega_m$), which means that the quality factor would be inversely proportional to frequency if this was the dominant loss, again favoring stronger Brillouin interaction at lower frequencies.
However, it is clear from our modeling that viscosity is not the limiting factor, as shown in Fig. \ref{Fig:qfactors}(b).
Leakage through the cladding varies from mode to mode and the profiles in Fig.~\ref{Fig:lockin} illustrate that some modes are more confined to the core than others.
In the modeling, this was accounted for using a mechanical perfect matched layer at the outer silica cladding surface (see section S1.B of the Supplementary Material for details).
Leakage is dominant above 400~MHz, as shown in Fig. \ref{Fig:qfactors}(b).

The main limitation for the low-frequency flexural modes arises from the effect of squeezed film air damping phenomenon, by which a micro-vibrating membrane transfers part of its energy to surrounding gas molecules\cite{bao_squeeze_2007}.
Since there are many membranes vibrating in phase inside the \gls{PCF} structure, this mechanism can be significant\cite{koehler2013effects}.
We used Bao's model\cite{bao_squeeze_2007} to estimate this contribution.
From the results in Fig. \ref{Fig:qfactors}(b), it is clear that squeezed film air damping is strongest for low order flexural modes, limiting their quality factors.
For the 90~MHz peak, this mechanism lowers the quality factor from 1800 to 33, or in terms of linewidth broadening, it is increased from tens of kHz up to 3~MHz, consistent with the experimental value of approximately 5~MHz.
At high frequencies, the calculated broadening due to squeezed film damping is negligible, and the obtained value of 35~MHz for the 2.13~GHz peak is consistent with the experimental value of 42~MHz. Another potential linewidth broadening factor not considered in the analysis here is the geometry non-uniformity along the fiber length.
According to our simulations, a 5\% geometry scaling results in 18\% and 8\% Brillouin frequency variation for 90~MHz and 2.13~GHz peaks, respectively.
In other words, this form of inhomogeneous broadening due to geometry fluctuation might affect low frequency flexural modes more strongly than higher ones. When all damping effects are considered, the simulated gain values drop to 6.9~$\textrm{W}^{-1}\textrm{km}^{-1}$, 20.3~$\textrm{W}^{-1}\textrm{km}^{-1}$ and 19.6~$\textrm{W}^{-1}\textrm{km}^{-1}$, consistent with the experimental values from Table \ref{tab:table1}. From a practical point of view, one could envision engineering structures that better confine mechanical modes to the core (i.e. reduce leakage), and rely on larger structures suspended-core fibers\cite{dong_highly_2008} to minimize squeezed film damping. In this ultimate scenario, the Brillouin gain for the 90~MHz flexural mode would be limited by viscosity as a fundamental mechanism, and could potentially reach values higher than 1000~$\textrm{W}^{-1}\textrm{km}^{-1}$. 

In conclusion, we demonstrated intermodal forward Brillouin scattering in \gls{PCF} between the HE$_{11}$ and HE$_{21}$ modes over a wide range of frequencies (90~MHz - 2.5~GHz), with a maximum gain coefficient of 21.4~$\textrm{W}^{-1}\textrm{km}^{-1}$. Symmetry arguments support the observation of polarization dependent and polarization independent intermodal Brillouin interaction, and different damping mechanisms dictate the relative strength of \gls{FBS} involving mechanical modes at low and high frequencies. This work opens the path to future engineering of \gls{PCF} structures to enhance the interaction of optical modes of high order through the control of flexural mechanical modes and their dissipation mechanisms, leading ultimately to novel physical phenomena and highly efficient devices.


\section*{Data Availability}
The data that support the findings of this study are available from the corresponding author upon reasonable request.

\begin{acknowledgments}
This research was funded by the São Paulo State Research Foundation (FAPESP) through Grant Nos. 08/57857-2, 13/20180-3, 18/15577-5, and 18/25339-4 and by the National Council for Scientific and Technological Development (CNPq) (Grant No. 574017/2008-9). This study was also partially funded by the Coordenação de Aperfeiçoamento de Pessoal de Nível Superior—Brasil (CAPES) - Finance Code 001. We would like to acknowledge support from Michael H. Frosz and Philip St. J. Russell from the Max Planck Institute for the Science of Light (Erlangen, Germany) in providing the photonic crystal fiber used in this study.

\end{acknowledgments}

\section*{References}
\bibliography{main_complete}

\end{document}


\title{Supplementary material: Intermodal Brillouin scattering in solid-core photonic crystal fibers    }

\author{Paulo F. Jarschel}
  \email{jarschel@ifi.unicamp.br}
  \affiliation{Photonics Research Center, University of Campinas, Campinas 13083-859, SP, Brazil\looseness=-1}
  \affiliation{Gleb Wataghin Physics Institute, University of Campinas, Campinas 13083-859, SP, Brazil\looseness=-1}
\author{Erick Lamilla}
  \affiliation{Photonics Research Center, University of Campinas, Campinas 13083-859, SP, Brazil\looseness=-1}
  \affiliation{Gleb Wataghin Physics Institute, University of Campinas, Campinas 13083-859, SP, Brazil\looseness=-1}
\author{Yovanny A. V. Espinel}
  \affiliation{Photonics Research Center, University of Campinas, Campinas 13083-859, SP, Brazil\looseness=-1}
  \affiliation{Gleb Wataghin Physics Institute, University of Campinas, Campinas 13083-859, SP, Brazil\looseness=-1}
\author{Ivan Aldaya}
  \affiliation{Campus of São João da Boa Vista, State University of São Paulo, São João da Boa Vista 13876-750, Brazil\looseness=-1}
  \affiliation{Photonics Research Center, University of Campinas, Campinas 13083-859, SP, Brazil\looseness=-1}
\author{Julian L. Pita}
  \affiliation{Photonics Research Center, University of Campinas, Campinas 13083-859, SP, Brazil\looseness=-1}
  \affiliation{Gleb Wataghin Physics Institute, University of Campinas, Campinas 13083-859, SP, Brazil\looseness=-1}
  \affiliation{School of Electrical Engineering, University of Campinas, SP, Campinas 13083-852, Brazil\looseness=-1}
\author{Andres Gil-Molina}
  \affiliation{Gleb Wataghin Physics Institute, University of Campinas, Campinas 13083-859, SP, Brazil\looseness=-1}
  \affiliation{School of Electrical Engineering, University of Campinas, SP, Campinas 13083-852, Brazil\looseness=-1}
  
  

\author{Gustavo S. Wiederhecker}
  \affiliation{Photonics Research Center, University of Campinas, Campinas 13083-859, SP, Brazil\looseness=-1}
  \affiliation{Gleb Wataghin Physics Institute, University of Campinas, Campinas 13083-859, SP, Brazil\looseness=-1}
\author{Paulo Dainese}
  \affiliation{Photonics Research Center, University of Campinas, Campinas 13083-859, SP, Brazil\looseness=-1}
  \affiliation{Gleb Wataghin Physics Institute, University of Campinas, Campinas 13083-859, SP, Brazil\looseness=-1}
  \affiliation{Corning Research \& Development Corporation, One Science Drive, Corning, New York 14830, USA\looseness=-1}

\date{\today}

\newcommand\blfootnote[1]{
  \begingroup
  \renewcommand\thefootnote{}\footnote{#1}
  \addtocounter{footnote}{-1}
  \endgroup
}
\blfootnote{\begin{addmargin}[-0.1em]{0em}Paulo F. Jarschel and Erick Lamilla contributed equally to this work.\end{addmargin}}

\maketitle
\tableofcontents

\section{\label{section:sim}Detailed simulation description}

\begin{figure*}[t!]
    \includegraphics[scale=1.0]{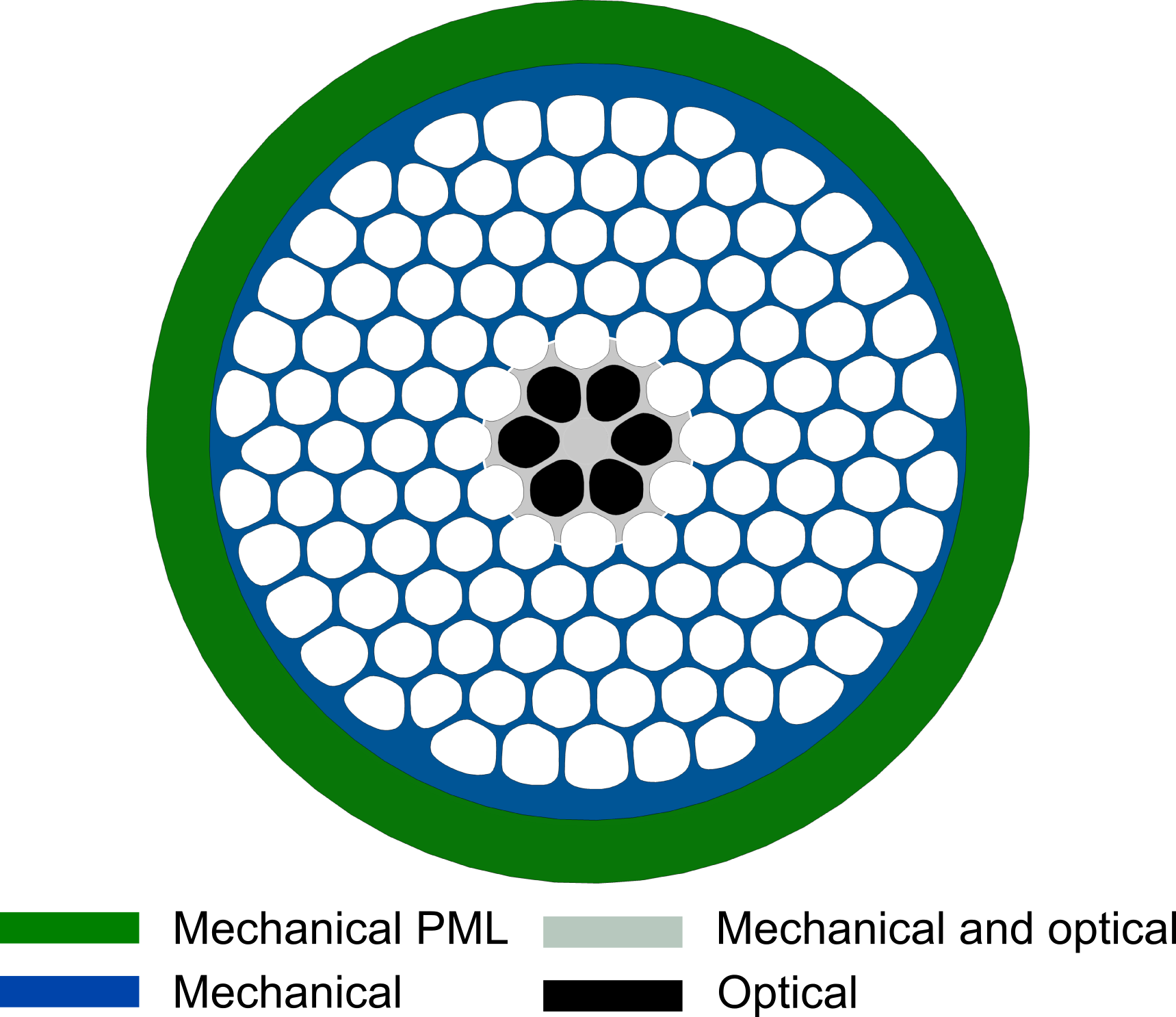}
    \caption{\label{fig:fem} Fiber cross-section extracted from the SEM image, with colors indicating the domains used to model optical and mechanical modes. The mechanical wave equation was solved in the grey and blue regions, while the electromagnetic wave equation was solved in the gray and black regions. A mechanical perfectly matched layer (PML) boundary condition was applied in the green region, and a perfect electric conductor (PEC) boundary condition was applied in the external boundary of the grey region. The core radius is $r_\text{c}\approx 1.7$~\textmu m, and the separation between holes is approximately 4.2~\textmu m. The PML thickness is 4.2~\textmu m.}
    \label{fig:sim_domain}
\end{figure*}

In this section, we detail the most important aspects of modeling forward stimulated Brillouin scattering (FSBS) in a solid core \gls{SC-PCF}. These results were used to generate Fig.~3 of the main text. In order to find both optical and mechanical modes, the full-vectorial wave equations  were independently solved by the finite elements method (FEM). Specifically, we employed the  \textit{Electromagnetic Waves} and \textit{Solid Mechanics} modules of \cmphys software (version 5.4). The computational domains used in our simulations are shown in \cref{fig:fem}. All simulations were carried out at a fixed optical wavelength ($\lambda=1.55$~\textmu m), and we used the material properties listed in Table \ref{tab:material_prop}.

\begin{table}[h]
\begin{tabular}{ c | c  } 

  Physical property & Value \\ 
\toprule
  Refractive index $n$ & 1.444 \\ 
 \hline
 Density $\rho$ [kg/m$^3$] & 2203 \\ 
 \hline
 Young modulus $Y$ [GPa] &73 \\ 
  \hline
 Poisson ratio $\nu_P$  & 0.17 \\
   \hline
 Viscosisty tensor  $(\eta_{11},\eta_{44}) $ [Pa.s] & $(2.79,1.04)\times 10^{-3}$ \\ 
  \hline
Photoelastic tensor  $(p_{11},p_{12},p_{44})$ & (0.121,0.27,-0.0745)  \\ 
\end{tabular}
\caption{Material properties used in the simulations.}
\label{tab:material_prop}
\end{table}

\subsection{Optical modes\label{sub:opt_modes}}






%

\begin{figure*}[ht!]
\includegraphics[scale=1]{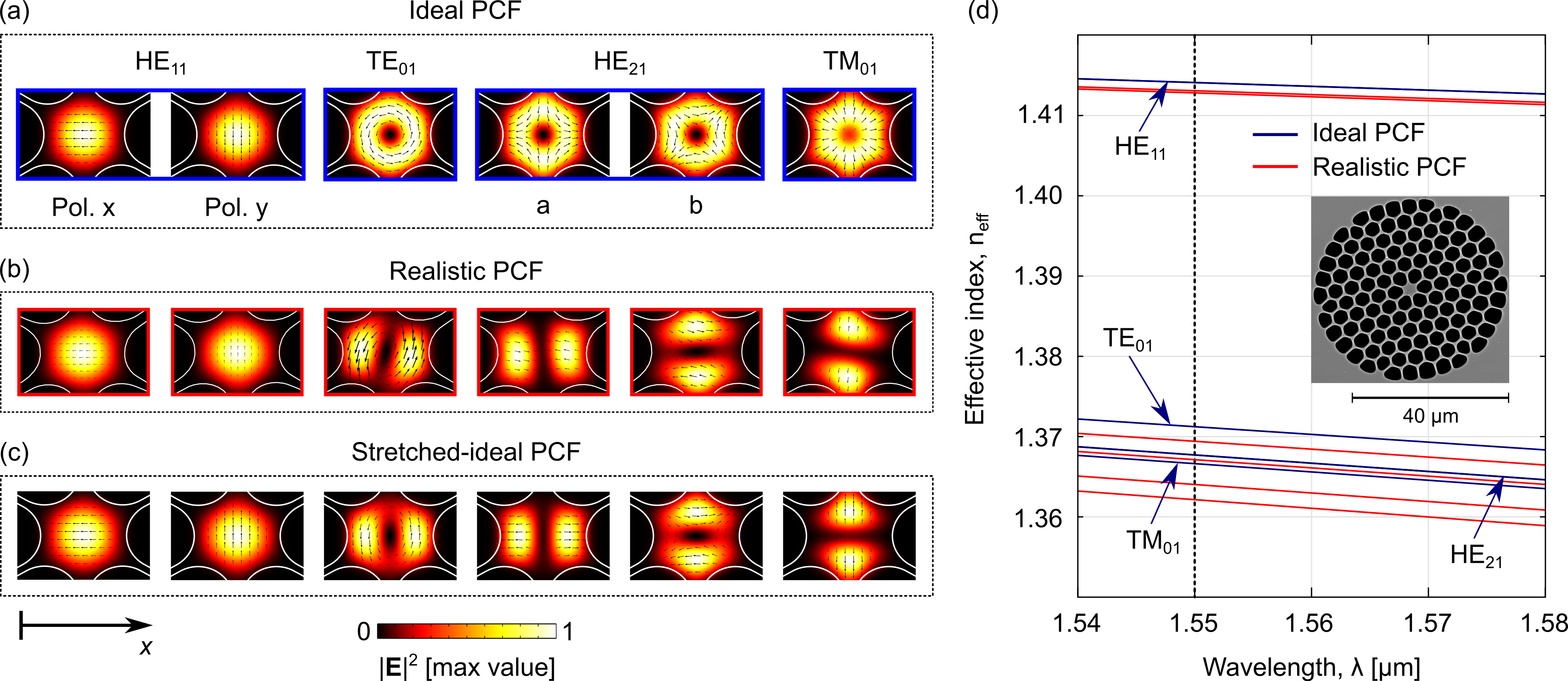}
\caption{\label{Fig:optmodes} Numerical analysis of the optical modes supported by the \gls{SC-PCF}: 
(a,b,c) represent the spatial profiles (squared electric field) at $\lambda=1.55$~\textmu m for the ideal, realistic, and stretched-ideal PCF geometries, respectively. (d) Modal dispersion (blue lines) of the HE$_{11}$, TE$_{01}$, HE$_{21}$ and TM$_{01}$, considering an ideal PCF with mean separation of 4.2~\textmu m between two adjacent holes and $r_\text{c}=1.7$~\textmu m. 
Red lines correspond to the dispersion relation curves of the same modes for a realistic PCF.} 
\end{figure*}

All simulations shown in the main text were carried out using the realistic PCF geometry shown in \cref{fig:sim_domain}, which was obtained from the SEM cross-section image. However, in order to precisely identify all the optical modes observed in the experiment, we compared the realistic geometry solutions with two idealized ones: (i)~an ideal triangular-lattice PCF structure, and (ii)~a stretched version of this ideal PCF to account for the asymmetry present in the actual structure (6\% stretching along the $x$-axis). While in the ideal PCF, the higher order modes shown in \cref{Fig:optmodes}(a) display the usual donut-shaped intensity profile, this is not the case for the actual fiber modes, shown in \cref{Fig:optmodes}(b). Despite the subtle differences between the ideal and actual fiber geometries, the optical modes of the stretched-ideal geometry,  shown in \cref{Fig:optmodes}(c), clearly capture most features of the actual fiber. Notably, the donut-like shape of the higher-order modes is lost as a result of lifting the degeneracy of the two polarization states of the HE$_{21}$ modes. Therefore, as far as the optical mode is concerned, the impact of geometry distortion is rather predictable. Mode shapes provide a first guidance to help identify which modes are excited experimentally, along with modal group delay. In \cref{Fig:optmodes}(d) we show the effective index ($n_\text{eff}$) for various modes, from which we can calculate the corresponding group index ($n_g=n_\text{eff}-\lambda \partial_\lambda n_\text{eff}$) and the differential group delay $\tau$. The results for the realistic structure are shown in Table \ref{tab:groupdelays}, where the differential group delays are calculated relative to the fundamental mode HE$_{11x}$ ($\tau=\Delta n_\text{g}L/c$, using the fiber length, $L=30$~m). In \cref{sec:optical_mode_exc} it is shown how these distinct group-indexes and mode profiles enables the identification of the optical modes excited in our experiment.

\begin{table}
\begin{tabular}{ c | c |c |c |c  |c|c} 

  & HE$_{11x}$ & HE$_{11y}$ & TE$_{01}$ & HE$_{21\text{a}}$ & HE$_{21\text{b}}$ & TM$_{01}$\\ 
 \toprule
  $n_g$ & 1.4687 & 1.4691 & 1.5025 & 1.5063 & 1.5084 & 1.5096\\
 \hline
  $\tau$ (ns) & - & 0.04 & 3.38 & 3.77 & 3.97 & 4.10\\
 \hline
  $\tau/L$ (ps/m) & - & 1.3 & 113 & 126 & 132 & 137\\
 
\end{tabular}
\caption{Simulated group index, and differential group delays for various modes (relative to the HE$_{11x}$ mode).}
\label{tab:groupdelays}
\end{table}





\subsection{\label{sec:mech_modes} Mechanical modes}

\begin{figure*}[t!]
\includegraphics[scale=1.0]{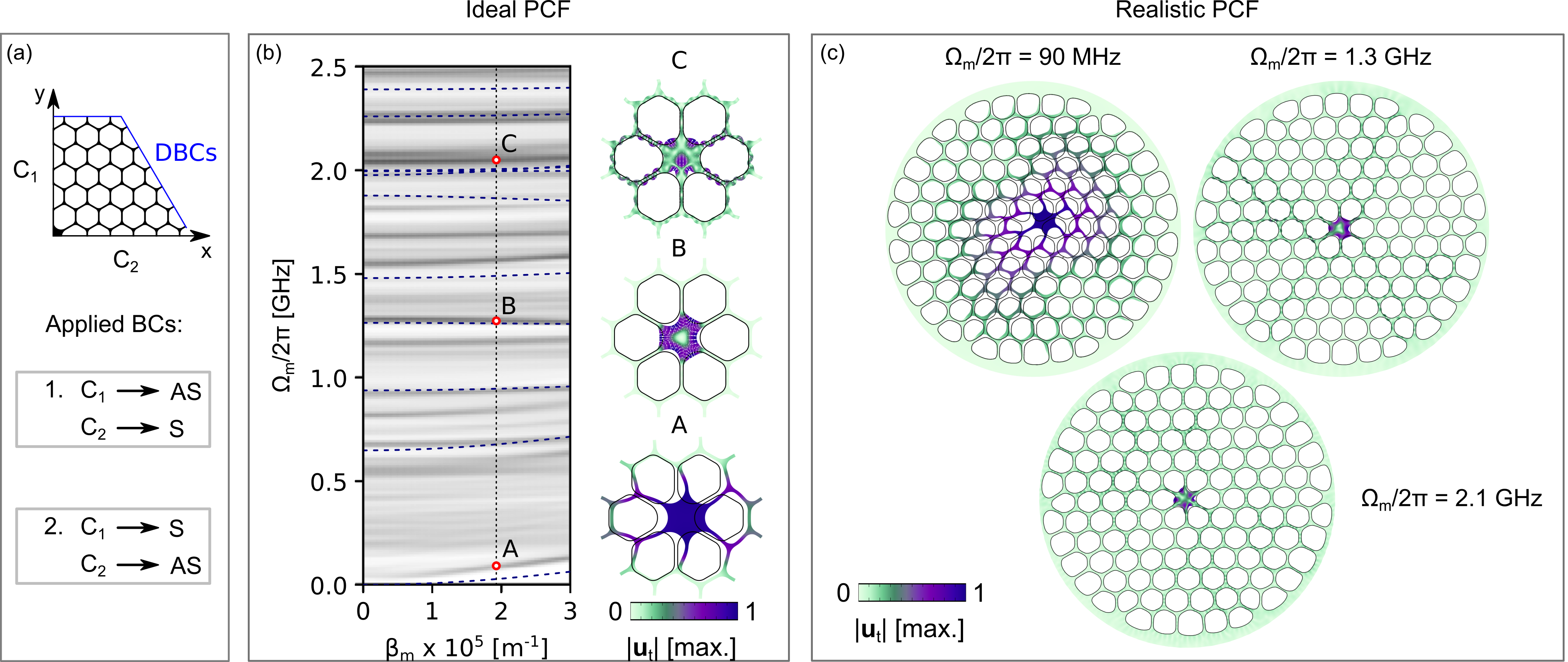}
\caption{\label{Fig:mechmodes} Numerical analysis of PCF mechanical modes. (a) schematic of the quarter structure and boundary conditions used to model the ideal PCF. We applied Dirichlet Boundary Conditions (DBCs) in the last ring, $\mathbf{u}=0$, and symmetric (S, $\mathbf{u}\cdot\hat{n}=0$), or antisymmetric (AS, $\mathbf{u}\cdot\hat{t}=0$) boundary conditions along the $x$-axis (C$_2$) and the $y$-axis (C$_1$) to model the flexural modes. (b) dispersion relation curves for all modes from the ideal PCF (grey density map) and dispersion curves for flexural modes in a rod waveguide (dashed blue lines). The dashed vertical line at $\beta_\text{m}=1.88\times 10^5$~m$^{-1}$ determines the phase-matching for intermodal Brillouin scattering involving HE$_{11x}$ and HE$_{21a}$ modes. Red circles correspond to the main  mechanical modes observed in the experiments (profiles shown on the right inset). (c) Correspondent mode profiles at $\Omega_\text{m}/2\pi=90$~MHz, $\Omega_\text{m}/2\pi=1.3$~GHz and at $\Omega_\text{m}/2\pi=2.1$~GHz, now obtained in the realistic PCF structure. Color scale represents the magnitude of the transverse displacement vector.}


\end{figure*}

Similarly to the optical modes, although the actual PCF geometry of  \cref{fig:sim_domain} was used in all simulations, exploring an idealized structure is helpful in identifying the mechanical mode families observed in the experiment. Due to specular symmetry with respect to both Cartesian axes, only a quarter of the full structure is used, which not only spares computational resources but also allows using boundary conditions that filters specific mode families. We applied Dirichlet Boundary Conditions (DBCs) in the last ring, $\mathbf{u}=0$, and symmetric (S, $\mathbf{u}\cdot\hat{n}=0$), or antisymmetric (AS, $\mathbf{u}\cdot\hat{t}=0$) boundary conditions along the $x$-axis (C$_2$) and the $y$-axis (C$_1$) to model the deformation of the flexural modes, as shown in \cref{Fig:mechmodes}(a).
In \cref{Fig:mechmodes}(b) we show the mechanical dispersion relation of the flexural modes in the ideal PCF, where mechanical eigenfrequencies were sought up to 2.5~GHz. Due to the large spectral density of mechanical modes, we choose to represent them as a density plot in \cref{Fig:mechmodes}(b).
The color scale represents the fraction of mechanical energy that is concentrated in the glass core (black 100\%, white 0\%), indicating which modes might interact more strongly with the optical modes. We can see that the dispersion relation curves do not vary significantly in this range. These characteristics are also reflected in the realistic PCF.
In the same plot, we show the dispersion relation for the flexural modes in a circular rod waveguide with equivalent dimensions (same core radius, $r=1.7$~\textmu m). Several modes in the PCF have dispersion and profile similar to those in the rod. For example, mechanical displacement profiles at $\beta_{\text{m}}=1.88\times 10^5$~m$^{-1}$ are illustrated in \cref{Fig:mechmodes}(b) for three modes A, B and C in the ideal PCF, and closely resemble their corresponding modes in the rod. Furthermore, these modes also have their correspondents in the actual PCF. For example, \cref{Fig:mechmodes}(c) shows the flexural-like modes at $\sim$90~MHz, $\sim$1.3~GHz and $\sim$2.1~GHz. Note that for the low frequency mode, its mechanical displacement field is distributed throughout the core and cladding, whereas the high frequency mode is mainly concentrated in the core region. These results suggest a stronger influence of low-frequency mechanical modes upon the optomechanical coupling process.

\subsubsection{\label{sec:mech_dissipation}Mechanical dissipation}




\paragraph{Leakage loss mechanism:} 
 
Leakage loss accounts for the energy escaping towards the solid silica region surrounding the PCF structure. To quantify its contribution, we included a perfectly matched layer (PML)~\cite{jin2015finite,comsol54smm} wrapping the photonic crystal, as shown in Fig.~\ref{fig:fem}, and we calculated the energy flowing on it. To implement a PML in COMSOL\textsuperscript{\tiny\textregistered}, the complex coordinate stretching function used is \mbox{$f(\rho)=\lambda_\text{t}s\rho^P(1-i)$}~\cite{comsol54smm}, where $\lambda_\text{t}$ is the \textit{typical wavelength} of the transverse component of the mechanical wave deformation ($\boldsymbol{u}$), $s$ is the \textit{scaling factor} that tunes the wave amplitude arriving at the edge of the PML, and $P$ is the \textit{scaling curvature parameter} that controls how strong is the amplitude decay inside the PML. The PML parameters used were: $s=0.25$ and $P=2.0$. We employed \mbox{$\lambda_\text{t}=2\pi[(\Omega_\text{m}'/V_\text{s})^2-\beta_\text{m}^2]^{-1/2}$}, with $V_\text{s}=3766$~m/s as the shear bulk velocity in silica, and \mbox{$\beta_\text{m}=2\pi(n_\text{eff,p}-n_\text{eff,S})/\lambda=1.86\times 10^5$~1/m}, from momentum conservation. The effective indexes $n_\text{eff,p}=1.413067$ and $n_\text{eff,S}=1.367107$ are the calculated at $\lambda=1.55$~\textmu m for the pump (HE$_{11x}$) and Stokes (HE$_{21a}$), respectively. Finally, $\Omega_\text{m}'$ is a mechanical frequency initial guess near the eigenfrequencies being sought for.

\paragraph{Viscous damping}
Viscous damping (VD) was included in the PCF model as an additional stress term proportional to the elastic strain rate~\cite{auld1973acoustic}, $\mathbf{\sigma}_{\text{VD}}=\mathbf{\eta}:\frac{\partial\mathbf{S}}{\partial t}$, where $\mathbf{\eta}$ is known as the fourth-rank viscosity tensor. The mechanical wave equation is then written as:
\begin{equation}
\tilde{\nabla} \cdot(\mathbf{c}': \mathbf{S})=-\rho\Omega^{2} \mathbf{u},\label{eq:wave-eq-mech-vd-pml}
\end{equation}

\noindent with $\mathbf{c}'=\mathbf{c}+i\Omega\mathbf{\eta}$ and $c$ being the stiffness tensor. The tilde in the divergence operator above indicates the coordinates transformation that arises from the PML. As fused silica is an isotropic material, the stiffness tensor has only two independent coefficients, $c_{11}=78.5$~GPa and $c_{44}=31.2$~GPa~\cite{auld1973acoustic}, as well as the viscosity tensor, $\eta_{\text{11}}=0.00279$~Pa$\cdot$s and \mbox{$\eta_{\text{44}}=0.00104$~Pa$\cdot$s}~\cite{auld1973acoustic,Krischer1970}. This equation was solved in \cmphys \textit{Structural Mechanics module} using isotropic materials (details in Subsection~E) with the following parameters: Young's modulus, $E=c_{44}(3c_{11}-4c_{44})/(c_{11}-c_{44})=73$~GPa,  Poisson's coefficient, $\nu_P=(c_{11}-2c_{44})/[2(c_{11}-c_{44})]=0.17$, bulk viscosity coefficient~\cite{boyd2008nonlinear}, $\eta_{\text{b}}=2\eta_{44}/3+\eta_{11}=0.00348$~Pa$\cdot$s.

\paragraph{Squeeze-air film damping}
The presence of air in the micro-metric channels in the PCF structure leads to a loss mechanisms known as squeeze-air film damping (SFD)~\cite{BAO20073}. The \textit{energy transfer model} has been successfully applied to different micro-electro-mechanical systems (MEMS), including  slab-air-substrate systems and microstructured optical fibers with double-membrane~\cite{Koehler2013,BAO20073}. According to this model, the SFD linewidth in a slab-air-substrate system is given by: 
\begin{equation}
\Gamma_\text{SFD}=\frac{2}{(2 \pi)^{3 / 2} \rho}\left(\frac{S}{H h_0}\right) \sqrt{\frac{M_\text{m}}{R T}} p,\label{eq:linewidth-sfd}
\end{equation}

\noindent in which, $S$, $\rho$, and $H$ are the slab's length, mass density, and thickness, respectively.   
$h_0$ is the gap between the slab and the substrate, $M_\text{m}$ is the molar mass of the air, $R$ is the ideal gas constant, $T$ is the temperature, and $p$ denotes the pressure.
We heuristically extended this model to a bore by mapping each circular air-hole to the vibrating slab model described by \cref{eq:linewidth-sfd}. The slab length  is considered as the hole's perimeter ($S=2\pi r_c$), while the slab-substrate gap $h_0$ is mapped to an effective bore diameter, given by the area to perimeter ratio, $h_0=\pi r_\text{c}^2/S$. Such mapping is feasible for a purely radial motion within each bore, however, in a general mechanical mode of the PCF structure, not all holes are radially deformed (e.g., motion can be tangential to the boundary). Since we are considering only the squeeze film effect, we take this into account by  introducing an effective number of radially deformed holes given by:

\begin{equation}
N_\text{eff}=\frac{\int|\mathbf{u}_\text{t}\cdot\hat{n}|dl}{\max(|\mathbf{u}_\text{t}|)2\pi r_\text{c}},\label{eq:mhn}
\end{equation}

\noindent in which, $\mathbf{u}_\text{t}$ is the deformation in the PCF cross section of the mode, obtained by solving \cref{eq:wave-eq-mech-vd-pml}, and $r_\text{c}$ is the PCF core radius. The integral is performed across all silica-air boundaries defining each air hole.
With this, the SFD contribution to the linewidth of a given mechanical mode, $\gamma_\text{SFD}$, can be calculated combining \cref{eq:linewidth-sfd} and \cref{eq:mhn}:  
\begin{equation}
    \gamma_\text{SFD}=\Gamma_\text{SFD}\cdot N_\text{eff}.\label{eq:gamma-SFD}
\end{equation}

In our simulations we employed:
$S=2\pi r_\text{c}$, $H=0.3$~\textmu m (wall thickness separating any two holes), $h_0=\pi r_\text{c}^2/S=0.85$~\textmu m, $\rho=2203$~kg~m$^{-3}$ from Auld (1973)~\cite{auld1973acoustic}, $M_\text{m}=29$~kg~mol$^{-1}$ from Lide (2003)~\cite{lide2003crc},
$p = 10^5$ Pa,
and $T$=300 K.

\paragraph{Total mechanical Q-factor and linewidth}

We first solved \cref{eq:wave-eq-mech-vd-pml} in order to find the Q-factor from the combined leaky and viscous damping mechanisms, $Q_\text{L+VD}=\Re{(\Omega_\text{m})}/[2\Im{(\Omega_\text{m})}]$, where $\Re{(\Omega_\text{m})}$ and $\Im{(\Omega_\text{m})}$ are respectively the real and imaginary parts of the mechanical eigenfrequency, $\Omega_\text{m}$. Then, we extracted the SFD Q-factor from \cref{eq:gamma-SFD} as, $Q_\text{SFD}=\Re{(\Omega_\text{m})}/\gamma_\text{SFD}$. Finally, we obtain the total mechanical Q-factor, $Q_\text{m}=Q_\text{SFD}Q_\text{L+VD}/(Q_\text{SFD}+Q_\text{L+VD})$ and $\gamma_\text{m}=\Re{(\Omega_\text{m})}/Q_\text{m}$.

\subsection{Coupled-mode description of Brillouin interaction} \label{subsec:cme}
Using the formalism adopted by Wiederhecker, et al. (2019)~\cite{wiederhecker_brillouin_2019}, we consider the interaction between two optical modes corresponding to pump and Stokes waves oscillating at $\omega_p$ and $\omega_s$ with complex amplitudes $\tilde{a}_\text{p}$ and $\tilde{a}_\text{s}$, respectively, and $K$ mechanical modes with complex amplitudes ${\tilde{b}_1,\tilde{b}_2,...,\tilde{b}_K}$. Both optical and mechanical amplitudes are defined in such a way that $|\tilde{a}_\text{p}|^2$ and $|\tilde{a}_\text{s}|^2$ represent the number of photons per unit length, and $|\tilde{b}_k|^2$ the number of the phonons per unit length of the $k$-th mechanical mode.
In the presence of multiple mechanical modes, an effective complex gain coefficient can be derived. Assuming large mechanical damping ($v_\text{m} / \gamma_\text{m}\ll1$), steady-state regime ($\partial_{t} \rightarrow 0$), and neglect the Stokes and pump losses ($\alpha_\text{s}, \alpha_\text{p} \approx 0$), the coupled equations become:

\begin{subequations}\label{eq:CME-system-approx}
    \begin{equation}\label{eq:CME-pump-approx}
    v_\text{p} \partial_{z} \tilde{a}_\text{p}=-i \sum_{j=1}^K\tilde{g}_{0,j} \tilde{a}_\text{s} \tilde{b}_\text{j}
    \end{equation}
    \begin{equation}\label{eq:CME-stokes-approx}
    v_\text{s} \partial_{z} \tilde{a}_\text{s}=-i \sum_{j=1}^K\tilde{g}_{0,j}^{*} \tilde{b}_j^{*} \tilde{a}_\text{p}
    \end{equation}
    \begin{equation}\label{eq:CME-mech-approx}
    \tilde{b}_k=\frac{-i \tilde{g}_{0,k}^{*}\tilde{a}_\text{s}^{*}\tilde{a}_\text{p}}{\gamma_{\text{m},k}/2+i\Delta_{\text{m},k}}.
    \end{equation}
\end{subequations}
\noindent where $\partial_z$ is the partial derivative with respect to the spatial coordinate, and $v_\text{p}$, $v_\text{s}$, and $v_{\text{m},k}$ represent the group velocity of the pump, Stokes and the $k$-th mechanical mode, respectively. 
$(\cdot)^*$ denotes the complex conjugate operation, 
\mbox{$\Delta_{\text{m},k}=\Omega-\Re{(\Omega_{\text{m},k})}$}, in which $\Omega = \omega_\text{p}-\omega_\text{s}$. $\tilde{g}_{0,k}$ is the waveguide optomechanical coupling rate [$\sqrt{m}/s$], whose modulus is related to the Brillouin gain coefficient, $G_{\text{B},k}$~\cite{PhysRevA.93.053828,wiederhecker_brillouin_2019}. In the undepleted regime ($\partial_{z} \tilde{a}_\text{p} \approx 0$), the evolution of the Stokes signal is given by:

\begin{equation}\label{eq:CME-stokes-approx2}
    v_\text{s} \partial_{z} \tilde{a}_\text{s}= \sum_{j=1}^K
    \left[\frac{|\tilde{g}_{0,j}|^2}{\gamma_{\text{m},j}/2-i\Delta_{\text{m},j}}
    \right] |\tilde{a}_\text{p}|^2\tilde{a}_\text{s}.
\end{equation}

\noindent Using the pump power defined as $P_\text{p}=\hbar \omega_\text{p} v_\text{p}\left|\tilde{a}_\text{p}\right|^{2}$, and Q-factor as $Q_{\text{m},k}=\Re{(\Omega_{\text{m},k})}/\gamma_{\text{m},k}$, we can express equation \ref{eq:CME-stokes-approx2} in terms of an effective complex Brillouin gain coefficient $G_B^{c}$:

\begin{subequations}\label{eq:pde-stokes}
    \begin{equation}\label{eq:pde-stokes-gen}
   \partial_{z} \tilde{a}_\text{s}=\frac{P_\text{p}(G_\text{B}^\text{c})^{*}}{2}
\tilde{a}_\text{s},
    \end{equation}
    \begin{equation}\label{eqgain-complex}
   G_\text{B}^\text{c}=\sum_{j=1}^{K} \mathscr{L}_{j}^\text{c}(\Omega) G_{\text{B}, j},
    \end{equation}
    \begin{equation}\label{eq:complex-linewidth}
    \mathscr{L}_{k}^\text{c}(\Omega)=\frac{\gamma_{\text{m},k}/2}{\gamma_{\text{m},k}/2+i\Delta_{\text{m},k}}.
    \end{equation}
\end{subequations}


%
%

Explicitly, the effective gain coefficient is can be written in terms of the moving boundary ($f_\text{mb}$) and photoelastic force densities ($f_\text{pe}$)\cite{wiederhecker_brillouin_2019}:

\begin{equation}
G_{\text{B},k}(\Omega)=Q_{\text{m},k} \frac{2 \omega_\text{p} (\Omega)}{\bar{m}_{\mathrm{eff},k} \Re{(\Omega_{\text{m},k})}^{2}}\left|\int f_{\mathrm{mb,k}} d l+\int f_{\mathrm{pe},k} d A\right|^{2}.\label{eq:gain}
\end{equation}

The MB integral must be evaluated across the perimeters of all holes in the first PCF ring (where the optical mode is large), whereas the PE integral must account for the whole cross-section area involving mechanical and optical modes, as is the grey region in \cref{fig:fem}. Finally, the Stokes power defined as $P_\text{s}=\hbar \omega_\text{s} v_\text{s}\left|\tilde{a}_\text{s}\right|^{2}$ is:

\begin{equation}
P_\text{s}=P_{\text{s}, 0} \exp \left(\Re\left[G_\text{B}^\text{c}\right] P_\text{p} L\right),\label{eq:pump-stokes-rel}
\end{equation}

\noindent in which $P_{\text{s},0}=\hbar \omega_\text{s} v_\text{s}\left|\tilde{a}_{\text{s},0}\right|^{2}$ is the initial Stokes optical power and $L$ is the fiber length. According to \cref{eq:pump-stokes-rel} the small signal gain is given by the real part of the complex function $G_\text{B}^\text{c}$.

\subsection{Numerical results} \label{subsec:cme}

\begin{figure}[htb!]
\includegraphics[scale=0.65]{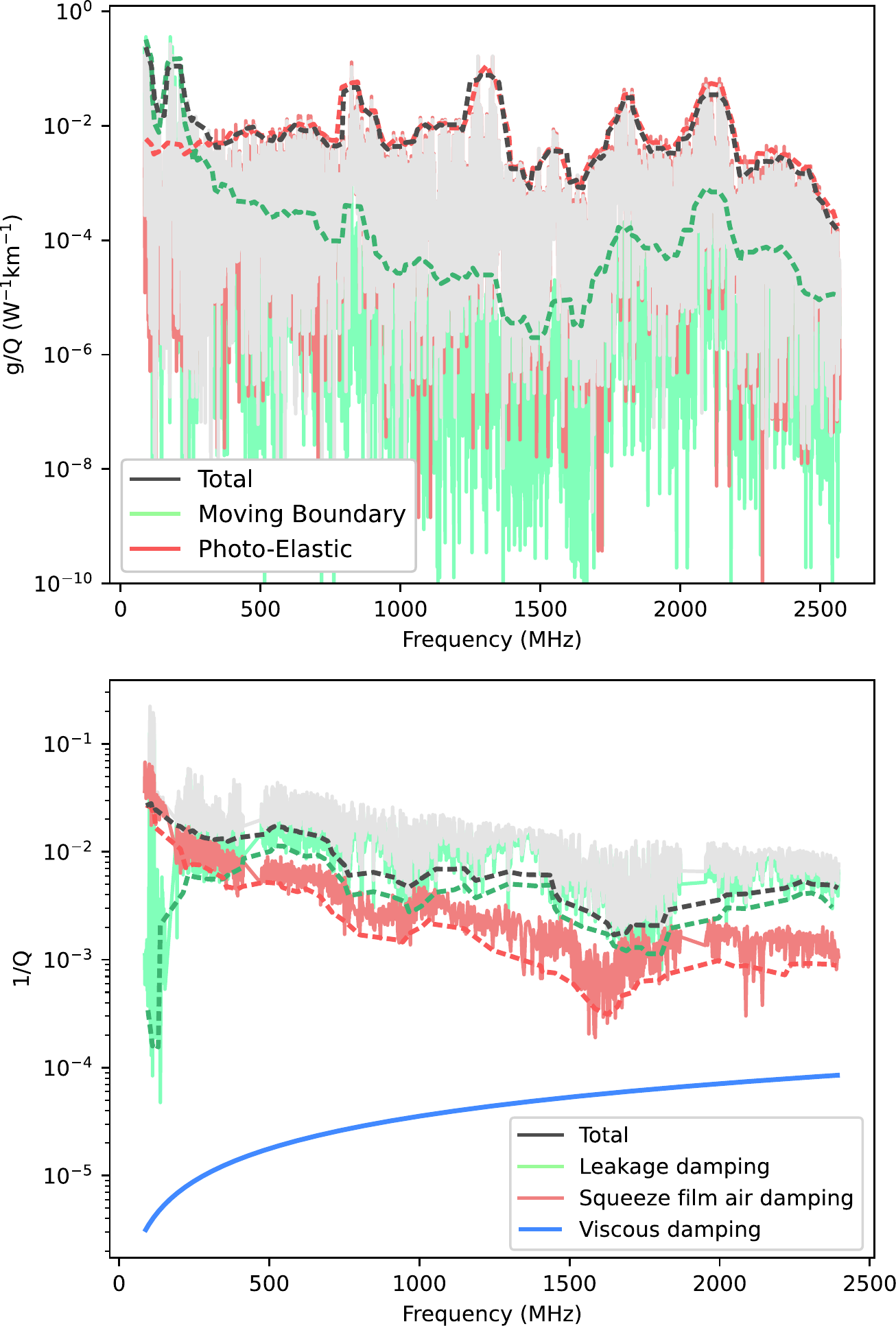}
\caption{\label{Fig:qfactors} (a) Calculated contributions to the net Brillouin gain (g/Q) from moving boundary and photo-elastic effects. (b) Calculated contributions from different mechanisms to the total mechanical damping, represented as the inverse of the quality factor ($Q^{-1}$). Dashed lines represent the envelope of the respective raw data.}
\end{figure}

Fig.~\ref{Fig:qfactors}, we plot separately $g/Q$, which represents the optomechanical coupling strength, and the inverse of $Q$, which represents the damping strength for all mechanical modes. The dashed curves represent envelop curves (highest $g/Q$ and lowest $1/Q$) as discussed in the main manuscript.

\section{\label{section:experiment}Experimental considerations}
\begin{figure*}[ht!]
\includegraphics[scale=0.9]{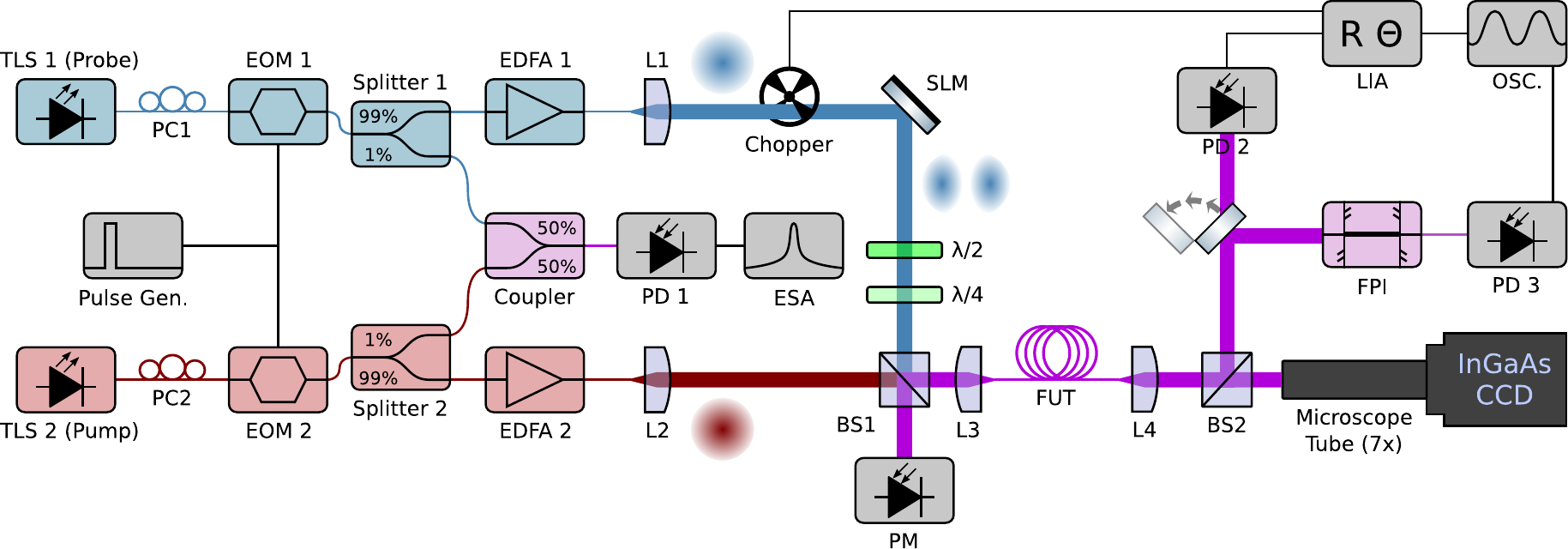}
\caption{\label{Fig:setup_complete}Experimental setup to characterize inter-modal power conversion in \gls{SC-PCF}s. TLS: tunable laser source; PC: polarization controller; EOM: electro-optic modulator; EDFA: Erbium-doped fiber amplifier; L: lens; PD: photodiode; ESA: electrical spectrum analyzer; SLM: spatial light modulator; $\lambda/2$, $\lambda/4$: wave plates; BS: beam splitter; PM: power meter; LIA: lock-in amplifier; OSC: oscilloscope; FPI: Fabry-Perot interferometer; FUT: fiber under test.}
\end{figure*}

Here we give a more in-depth description of how the distinct optical modes were excited and characterized (\cref{sec:optical_mode_exc}) and how the lock-in amplifier was employed to identify the active mechanical modes (\cref{sec:FBS_id}). The detailed version of our experimental setup is shown in \cref{Fig:setup_complete}.

\subsection{\label{sec:optical_mode_exc}Modal excitation and characterization}

\begin{figure}[!b]
\includegraphics[scale=1]{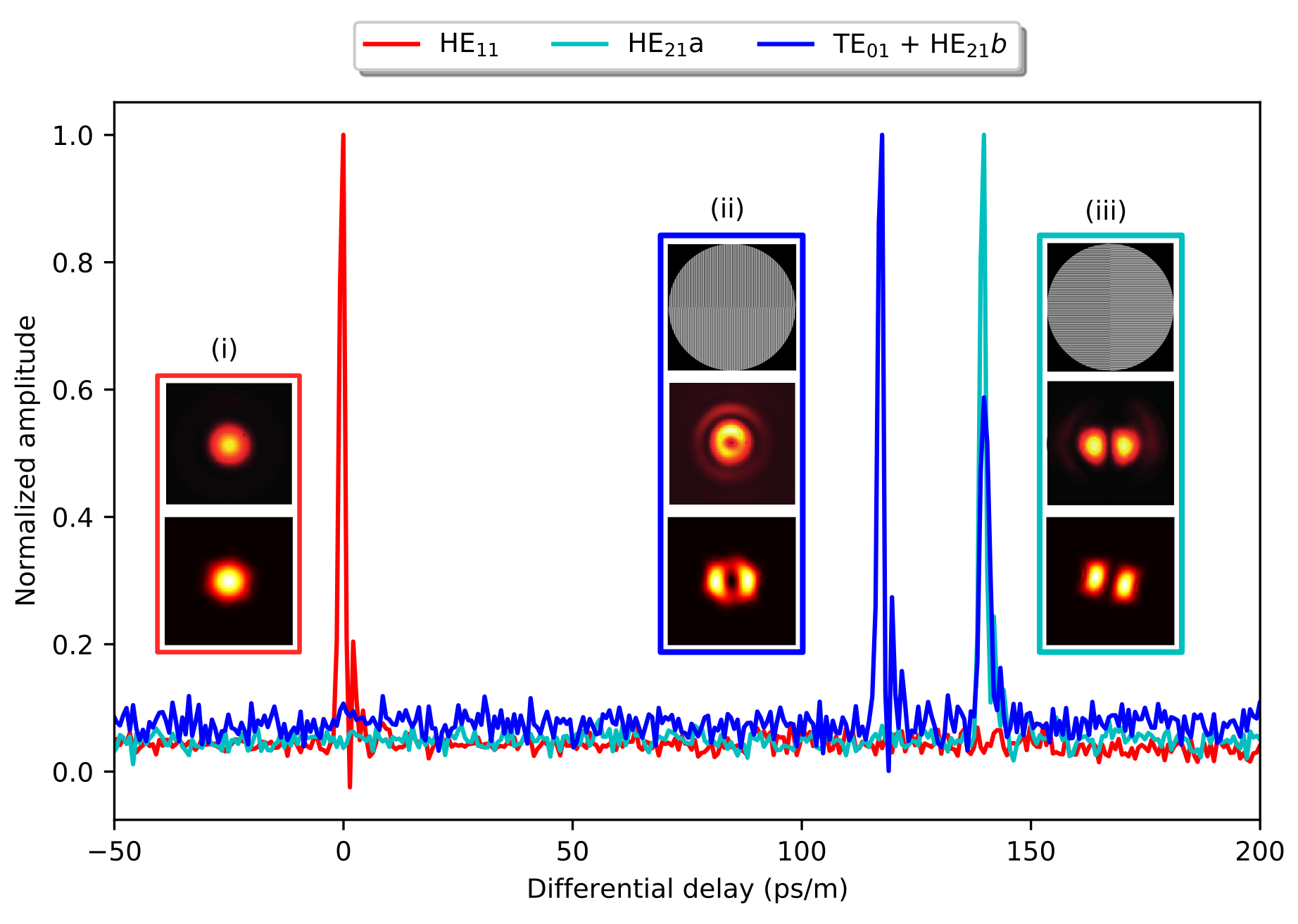}
\caption{\label{Fig:excitation} Output pulses for different excitation conditions. 
For each curve, the inset shows the intensity profiles captured by the InGaAs camera and simulated using finite-element method. 
For higher-order modes, the phase mask employed to excite the modes are also included.}
\end{figure}

In order to excite the HE$_{21}$a mode, we used the phase mask shown in the inset (ii) of Fig.~\ref{Fig:excitation}. This mask was built by multiplying a blazed grating and a $\pi$-difference phase mask. 
By doing so, we were able to use the first-order diffraction of the grating to convert the incident Gaussian beam into the desired higher order mode with improved beam quality.
The conversion efficiency of this approach, however, depends on the incident polarization, so a fiber polarization controller was included at the output of the fiber before conversion to free space. 
Additionally, with the aim of maximizing the matching between the output beam and the higher order mode of the fiber, the phase mask was rotated and shifted, whereas a $\lambda/2$-plate was included to minimize the polarization mismatch.  
It is important to note that during the realization of the experiments, the purity of the mode was periodically monitored by short pulse measurements as the ones shown in Fig.~\ref{Fig:excitation}, and by imaging the output beam with the InGaAs CCD camera. Note that for a phase mask with a $\pi$ along the horizontal axis (iii), the beam profile resembles well the HE$_{21}$a mode, and the measured group delay is 140 ps/m, which is in good agreement to the simulated value of 126 ps/m.



\subsection{\label{sec:FBS_id}FBS frequency identification}


\begin{figure}[!b]
\includegraphics[scale=0.9]{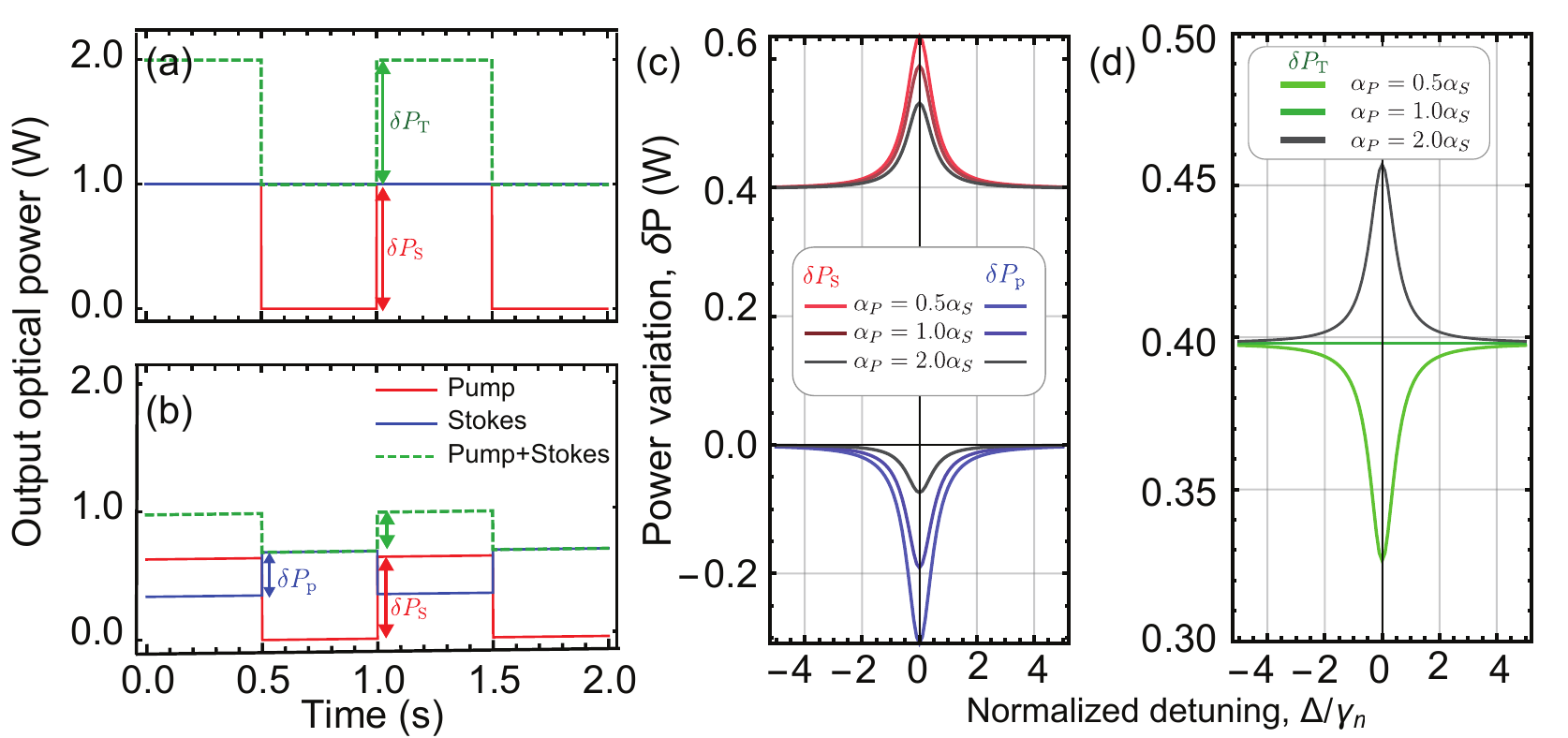}
\caption{\label{Fig:lockin} \textbf{(a)} Temporal profile of pump (blue) and Stokes (red) signals at the fiber input. The green-dashed trace represents the total power (sum) of both signals. \textbf{(b)} Same as \textbf{(a)}, but at the fiber output. The vertical arrows in \textbf{(a,b)} indicate the peak-to-peak amplitude of each signal. \textbf{(c)} Power variation dependence on the  pump-Stokes frequency detuning, measured at the output of the fiber ($L=39$~m) for different pump attenuation, clearly Similarly to the experiment, in all cases, the Stokes attenuation was set to $\alpha_S=0.15$~dB/m and $G_B=20$(W.km)$^{-1}$.}
\end{figure}

As indicated in \cref{Fig:setup_complete}, the probe laser (Stokes) is modulated using a chopper in order to enable a lock-in amplifier (LIA) detection scheme. This set the basis to build Figure 3 in the main text. It is not obvious, however, that signal resulting from this detection scheme should be related to the Stokes probe gain when the detuning matches a mechanical resonance. For example, for lossless optical modes, the Manley-Rowe relations\cite{wiederhecker_brillouin_2019} imply that the photons lost from the pump mode are transferred to the probe wave. Since the mechanical frequency is much smaller than the optical frequency, this is equivalent to state that the pump power depletion  must match the probe power gain. Since the signal generated at photodetector PD2 is proportional to the total power (pump + Stokes), no change would be observed on the LIA signal whether the lasers are tuned to a mechanical resonance or not. In other words, the net LIA signal in this lossless scenario would not be sensitive to the Brillouin gain. 

To investigate the connection between the LIA signal and the actual Stokes gain in a realistic lossy fiber, considering pump (fundamental mode) and Stokes (higher-order mode) signals with loss coefficients of 0.04~dB/m and 0.15~dB/m, we solved the steady-state \cref{eq:CME-system-approx} with a square-wave modulated probe signal. Since the bandwidth of the photodetectors PD2 or PM are much smaller than the pump-probe frequency detuning, their time-dependent output signal is simply given by the sum of pump and probe powers. At the fiber input, as it would be detected by photodetector PM, these signal are  depicted in \cref{Fig:lockin}(a). The simulated output pulses are shown in \cref{Fig:lockin}(b). At resonance, energy transfer occurs in the time intervals when the Stokes probe is on. However, the Stokes gain is no longer equal to the pump depletion, simply as a consequence of different attenuation rates ($\alpha_p \neq \alpha_S$). In other words, power flows from a low-loss channel (fundamental mode) into a high-loss channel (HE$_{11a}$). As a result, when the lasers are tuned to a mechanical resonance, the pump power variation ($\delta P_p$) is clearly different from the Stokes power variation($\delta P_S$). To confirm the importance of the loss imbalance for the lock-in detection scheme, we show in \cref{Fig:lockin}(c) how these power variations would depend on the detuning ($\Delta$) between the pump-Stokes beat-note ($\Omega=\omega_p-\omega_S$) and the mechanical frequency of a given mechanical mode, i.e., ($\Delta=\Omega-\Omega_m$ (normalized by the mechanical linewidth, $\gamma_m$), also, in \cref{Fig:lockin}(c) we show how different different values of pump and Stokes mode optical losses impact the LIA signal. For instance, if the two modes present exactly the same loss, the LIA signal would have a flat response. Also this model explains whether dips or peaks are to be observed in the LIA signal: if $\alpha_p>\alpha_S$, a peak would be observed, meaning that the Stokes gain contributes more to the signal, while if $\alpha_p<\alpha_S$, a dip is measured, indicating that the pump depletion is more significant at the output signal.
In order to obtain the traces in Fig. 3 of the main text,  the frequency detuning between the pump and Stokes was swept using the piezo-scan available in the probe-laser, so a temporal trace of the output power with well-defined dips could be obtained. It is worth noting that in order to isolate the power reduction due to intermodal power transfer, it is useful to measure the baseline in absence of FBS by blocking the pump wave. 

The frequency resolution of the employed method is limited by two factors. 
On the one hand, the use of pulses for both Stokes and pump poses a lower resolution boundary because the effective linewidth is broadened. 
As it is well known, longer pulses result in narrower linewidth but, unfortunately, it is not possible to arbitrarily increase the pulse length because longer pulses lead to lower peak power. However, the incident power limitation of the \gls{SLM} (200~mW) should also be considered.
Therefore, 80~ns pulses were adopted in our experiments since this value simultaneously allowed a relatively high peak input power of 2~W and an acceptable frequency resolution of 12.5~MHz. 
On the other hand, we should consider the intrinsic resolution of the LIA, which further depends on the pump-Stokes frequency sweeping speed and LIA integration time. These parameters should be chosen to optimize the trade-off between frequency resolution, and noise bandwidth.
In this way, for a fixed sweeping speed, a larger integration time leads to improved signal-to-noise ratio but poorer resolution and vice versa. 
To get the curves shown in Fig.~3(a) of the main manuscript, we configured the lock-in amplifier to have a sweeping speed of 300~MHz/s and an integration time of 30~ms, which results in a $\sim$9~MHz resolution. 
Therefore, to find the combined frequency resolution, we performed the convolution of a sinc function with a full-width at half-maximum (FWHM) of 12.5~MHz and a Lorentzian with a FWHM linewidth of 9~MHz, resulting in a total frequency resolution of 15.5~MHz.     

\subsection{\label{sec:gain_fiting} Brillouin gain characterization}

\begin{figure*}[hb!]
\includegraphics[width=16cm]{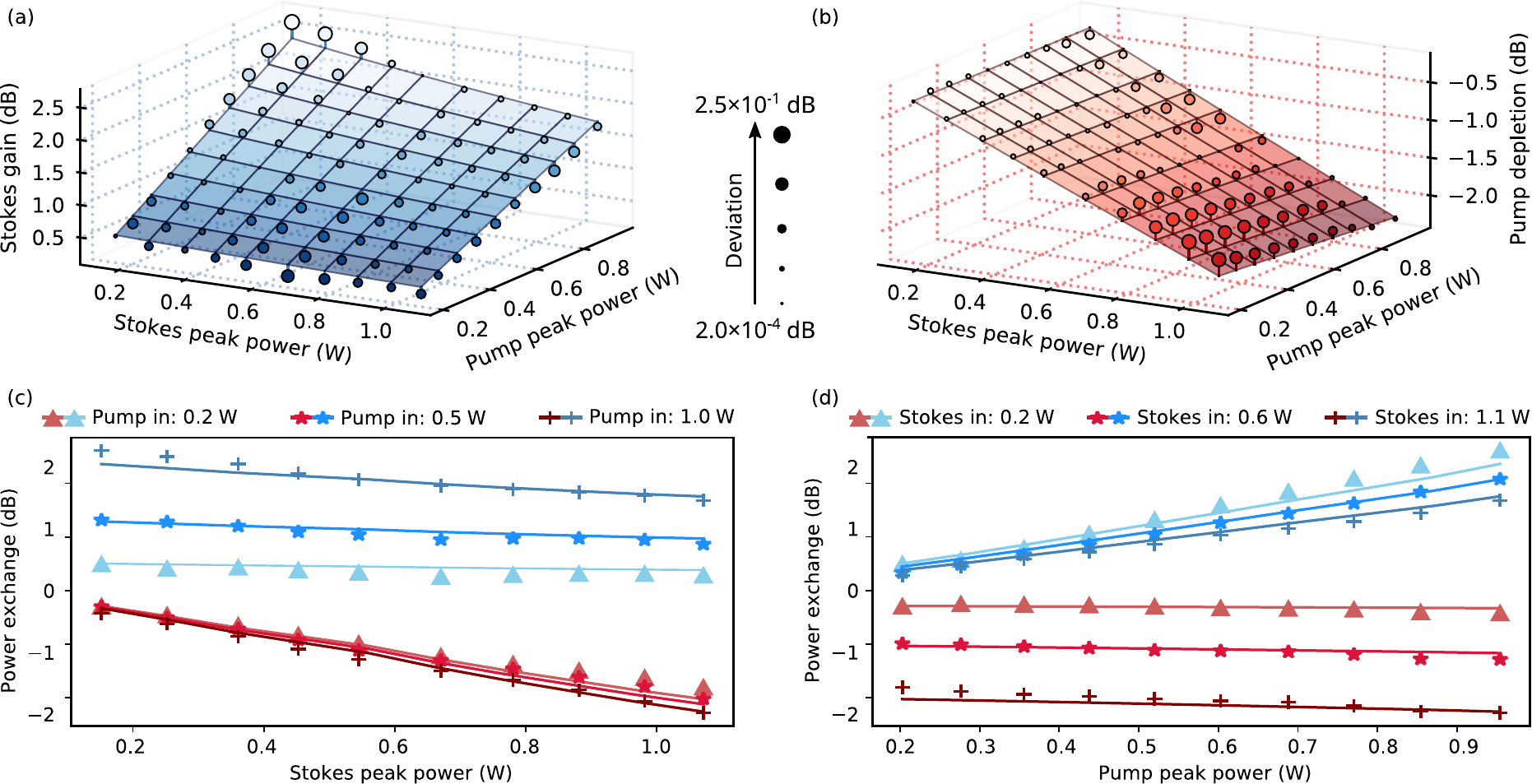}
\caption{\label{Fig:maps} Three-dimensional representation of the measured Stokes gain (a) and pump depletion (b), as a function of Stokes and pump input power. The circles correspond to the experimental data points, with their sizes proportional to the deviation from the surface obtained by fitting the experimental data to the model from equations (2) and (3) from the main manuscript. (c), (d) Two-dimensional cross-sections of the data from (a) and (b), respectively.}
\end{figure*}

To obtain the Brillouin gain coefficient, we measured several on/off gain (and depletion) points as a function of both Stokes and pump input powers, and the results are shown in Fig.~\ref{Fig:maps}(a) and Fig.~\ref{Fig:maps}(b), for the 2.13~GHz peak.
From the data points, we can see that the Stokes gain value reaches 2.6~dB for the Stokes and pump respective input powers of 0.95~W and 0.15~W, while the pump depletion reaches 2.8~dB for 0.95~W and 1.07~W.
By introducing loss into \cref{eq:CME-system-approx}, the power exchange is governed by the following coupled rate equations: 

\begin{equation}
    \label{eq:ratep}
    \frac{\text{d}P_p(z)}{\text{d}z}=-\alpha_pP_p(z)-gP_p(z)P_s(z)
\end{equation}
\begin{equation}
    \label{eq:rates}
    \frac{\text{d}P_s(z)}{\text{d}z}=-\alpha_sP_s(z)+gP_p(z)P_s(z),
\end{equation}

\noindent where $P_p$ and $P_s$ correspond to the pump and Stokes powers, respectively, $z$ is the position along the fiber, $\alpha$ corresponds to the attenuation coefficient for the different optical modes, and $g$ is the Brillouin gain coefficient. Equations (\ref{eq:ratep}) and (\ref{eq:rates}) were solved numerically using the measured attenuation coefficients for HE$_{11}$ (0.04~dB/km) and HE$_{21}$ (0.15~dB/km) modes, and leaving $g$ as the fitting parameter of the Least Squares method.



\bibliography{supplementary.bib}